\documentclass[twocolumn,showpacs,amsmath,amssymb,nofootinbib]{revtex4}% Physical Review D
\allowdisplaybreaks

\usepackage{graphicx}
\usepackage{amssymb}
\usepackage{epstopdf}
\usepackage{epsfig}
\allowdisplaybreaks

\begin{document}

\title{ Effect of sigma meson on the $D_1(2430) \to D\pi\pi$ decay}

\author{Masayasu Harada}

\author{Hironori Hoshino}

\author{Yong-Liang Ma}

\address{Department of Physics, Nagoya University, Nagoya, 464-8602, Japan.}
\date{\today}
%%%%%%%%%%%%%%%%%%%%%%%%%%%%%%%
\begin{abstract}

We study the effect of sigma meson on the $D_1(2430) \to D\pi\pi$
decay by constructing an effective Lagrangian preserving the chiral
symmetry and the heavy quark 
symmetry. The sigma meson is included through a linear sigma model, in
which both the $q\bar{q}$ and  $qq\bar{q}\bar{q}$ states are
incorporated 
respecting their different $U(1)_{ A}$ transformation 
properties. We first fit the sigma meson mass and $\sigma$-$\pi$-$\pi$
coupling constant to the $I=0$, $S$-wave $\pi$-$\pi$ scattering data. 
Then, we show how the differential decay width $d\Gamma( D_1 \to D
(\pi\pi)_{I = 0, L =0})/d m_{\pi\pi}$ depends on the 
quark structure of the sigma meson. We find that our study, combing
with the future data, can give a clue to understand the sigma meson
structure.  
\end{abstract}
%12.39.Hg Heavy quark effective theory
%14.40.Rt Exotic mesons
%11.30.Rd Chiral symmetries

\maketitle

\section{Introduction}

\label{sec:intro}

The lightest scalar meson ``sigma'' is an interesting object which 
may give a clue to understand some fundamental problems of QCD
such as the chiral symmetry structure, the origin of mass and so on.
 The mass spectrum of the light scalar meson nonet
including the sigma meson disfavors the $q\bar{q}$ picture but
prefers the $qq\bar{q}\bar{q}$ interpretation 
(see, e.g. Ref.~\cite{scalarreview,Nakamura:2010zzi} and references therein).

If in the nature there are both $q\bar{q}$ and $qq\bar{q}\bar{q}$
scalar states, they might mix to give the physical scalar mesons
(sometimes the glueball component is included). In the literature,
this superposition has been discussed widely~\cite{mixing}, 
but the structure of the
light scalar mesons remains an open question and deserves further
investigation. 

To investigate the quark contents of the light scalar mesons, we
should find some quantities that can distinguish different components,
two-quark component, four-quark component and glueball. 
In Ref.~\cite{Black:2000qq} it was pointed that, in a linear sigma
model expressing the 
spontaneous chiral symmetry breaking, the $q\bar{q}$ and
$qq\bar{q}\bar{q}$ bound states have different charges of $U(1)_{A}$
symmetry therefore one can use this $U(1)_{A}$ transformation property
to discriminate the  $q\bar{q}$ and $qq\bar{q}\bar{q}$ components of
the scalar mesons. The $U(1)_{A}$ symmetry is explicitly broken by
anomaly in QCD which causes a mixing between the $q\bar{q}$ states and
the $qq\bar{q}\bar{q}$ states~\cite{Fariborz:2005gm,Hooft:2008we,Fariborz:2008bd}. This contribution is included in the
model Lagrangian in such a way that the anomaly matching condition is
satisfied. Then, unlike the other terms which are invariant under the
$U(1)_{A}$ transformation, the $U(1)_{A}$ violating terms are
constrained by the anomaly. Another source of the mixing between
$q\bar{q}$ and $qq\bar{q}\bar{q}$ comes from the existence of both
the $q\bar{q}$ and $qq\bar{q}\bar{q}$ condensation~\cite{Fariborz:2005gm,Fariborz:2008bd}.

In this paper, we devote ourselves to study the sigma meson structure
in the heavy-light meson decay, specifically, the $D_1(2430) \to D
(\pi\pi)_{I = 0, L = 0}$ decay, based on the chiral partner structure
between $(D^\ast, D)$ and
$(D_1(2430),D_0^\ast(2400))$~\cite{Nowak:1992um,Bardeen:1993ae} (in
the following, we simply denote $D_1(2430)$ and $D_0^{\ast}(2400)$ as
$D_1$ and $D_0^{\ast}$, respectively). An interesting property of this
process is that there is only one light quark in the heavy-light meson
so that the axial transformation property is well determined and the
heavy-light meson couples only to the two-quark component of the
scalar meson. 

The paper is organized as follows: In sec.~\ref{sec:sigmamodel}, we
introduce an extended linear sigma model for three flavor
QCD. Section~\ref{sec:scattering} is devoted to study the $\pi$-$\pi$
scattering: We determine the $\sigma$-$\pi$-$\pi$ coupling and sigma
meson mass by fitting them to the data for the $I=0$, $S$-wave channel. 
In sec.~\ref{sec:heavyL}, we construct an effective Lagrangian for the
interaction among light mesons and heavy-light mesons. In sec.~\ref{sec:heavydecay}, after determining the relevant parameters in the heavy-light meson sector, we
show how the differential decay width $d\Gamma( D_1 \to D
(\pi\pi)_{I = 0, L =0})/d m_{\pi\pi}$ depend on the 
quark structure of the sigma meson. We give a summary and discussions in
sec.~\ref{sec:con}. In appendices we show some details of derivations of several
formulas.

\section{Linear Sigma model with two-quark and four-quark states}

\label{sec:sigmamodel}

In this section, we introduce a linear sigma model for three flavor
QCD in the low-energy region by including a $3 \times 3$ chiral nonet
field $ M $ representing the $q\bar{q}$ states and a $3 \times 3$
chiral nonet field $M^\prime$ standing for the $qq\bar{q}\bar{q}$
states. These two nonets have the same chiral $SU(3)_L \times SU(3)_R$
transformation property: 
\begin{eqnarray}
M \rightarrow g_L M g_R^\dag
\ ,
\quad
M^{\prime} & \rightarrow & g_L M^{\prime} g_R^\dag
\ ,
\label{eq:chirallight}
\end{eqnarray}
where $g_{L,R} \in SU(3)_{L,R}$. On the other hand, as pointed in
Ref.~\cite{Black:2000qq,Fariborz:2005gm}, they have the following different $U(1)_A$ transformation properties: 
\begin{eqnarray}
M & \rightarrow & e^{2i\alpha} M \,,
\;\;\;\;
M^\prime \rightarrow e^{-4i\alpha}
M^\prime , \,\label{eq:axiallight}
\end{eqnarray}
with $\alpha$ as the phase factor of the axial transformation. We
decompose $ M $ and $M^\prime$ as 
\begin{eqnarray}
M  = S + i \Phi
\ , \quad
M^{\prime} = S^{\prime} + i \Phi^{\prime}
\ ,
\end{eqnarray}
where $S$ and $S^{\prime}$ are the scalar meson matrices
and $\Phi$ and $\Phi^{\prime}$ are the pseudoscalar meson matrices.

In this paper we adopt the following extended linear sigma
model~\cite{Black:2000qq}: 
\begin{eqnarray}
{\cal L}_{\rm light} & = & \frac{1}{2}{\rm Tr}(\partial _\mu M \partial^\mu
M^{\dag}) + \frac{1}{2} {\rm Tr}(\partial _\mu M^\prime \partial^\mu M^{\prime
\, \dag}) \nonumber\\
& & {} - V_0(M,M^\prime) - V_{\rm SB} \ ,
\label{eq:lagrlight}
\end{eqnarray}
where the first two terms are the kinetic terms for the $q\bar{q}$ and
$qq\bar{q}\bar{q}$ states, $V_0(M,M^\prime)$ is the potential term
invariant under the chiral $SU(3)_L \times SU(3)_R$ transformation,
and $V_{\rm SB}$ stands for the explicit chiral symmetry breaking
terms due to the current quark masses. It should be noted that
$V_0(M,M^\prime)$ is decomposed as 
\begin{eqnarray}
V_0(M,M^\prime) = V_{\rm inv}(M,M^\prime) + V_{\eta}(M,M^\prime),
\end{eqnarray}
where $V_{\rm inv}$ is invariant under the $U(1)_A$ transformation
while $V_{\eta}$ violates the $U(1)_A$ symmetry explicitly due to the
anomaly. We do not specify the form of $V_0$ in our present analysis
but assume that this model allows for spontaneous chiral symmetry
breaking by a consistent choice of the parameters in $V_0$. Since the
$U(1)_{A}$ symmetry is explicitly broken by the anomaly, the form of
the $U(1)_{A}$ violating term $V_\eta$ is constrained by the anomaly
matching condition. 

The $U(1)_A$ symmetry breaking by anomaly causes a mixing between the
$q\bar{q}$ states and 
the $qq\bar{q}\bar{q}$ states, as pointed in
Refs.~\cite{Fariborz:2005gm,Hooft:2008we,Fariborz:2008bd}. 
%This is included in our model Lagrangian through the 
%$V_\eta(M,M')$ term in Eq.~(\ref{eq:lagrlight}). 
In addition,
% to this $U(1)_{A}$ violating term, 
the spontaneous chiral
symmetry breaking also generates a mixing between the $q\bar{q}$
states and $qq\bar{q}\bar{q}$
states~\cite{Fariborz:2005gm,Fariborz:2008bd}. As
a result, four physical iso-singlet scalar mesons are given as the
mixing states of two $q\bar{q}$ states and two $qq\bar{q}\bar{q}$ states through the mixing matrix $U_{f}$ as
\begin{eqnarray}
\left(
  \begin{array}{c}
     f_{p1} \\
     f_{p2} \\
     f_{p3} \\
     f_{p4} \\
  \end{array}
\right) = \left(
\begin{array}{llll}
(U_f)_{1a} & (U_f)_{1b} & (U_f)_{1c} & (U_f)_{1d} \\
(U_f)_{2a} & (U_f)_{2b} & (U_f)_{2c} & (U_f)_{2d} \\
(U_f)_{3a} & (U_f)_{3b} & (U_f)_{3c} & (U_f)_{3d} \\
(U_f)_{4a} & (U_f)_{4b} & (U_f)_{4c} & (U_f)_{4d} \\
\end{array}
\right) \left(
\begin{array}{c}
f_a \\
f_b \\
f_c \\
f_d \\
\end{array}
\right),\nonumber\\
\label{eq:mixings}
\end{eqnarray}
where $f_{p1}$, $\ldots$, $f_{p4}$ are the physical scalar mesons with mass ordering
$m_{f_{1}}\leq m_{f_{2}}\leq m_{f_{3}}\leq m_{f_{4}} $. In the
following, consistently with PDG notation, we use the notation
$\sigma$ for the lightest scalar meson $f_{p1}$. $f_a$ and $f_b$ are
the $q\bar{q}$ states with $f_a = (S_1^1+S_2^2)/\sqrt{2}$ and 
$f_b = S_3^3$ and $f_c$ and $f_d$ are the $qq\bar{q}\bar{q}$ states
with $f_c = ((S^{\, \prime})_1^1+(S^{\, \prime})_2^2)/\sqrt{2}$ and 
$f_b = (S^{\, \prime})_3^3$. 

Similarly, for the two iso-triplet pseudoscalar mesons, we have
\begin{eqnarray}
\left(
\begin{array}{l}
\pi_{p} \\
\pi_{p}^{\prime} \\
\end{array}
\right) & = &
% (U_\pi)_{ij} \left(
%\begin{array}{l}
%\pi  \\
%\pi^\prime  \\
%\end{array}
%\right)= 
\left(
\begin{array}{lr}
\cos \theta _\pi & -\sin \theta _\pi \\
\sin \theta _\pi & \cos \theta _\pi \\
\end{array}
\right) \left(
\begin{array}{l}
\pi  \\
\pi^\prime  \\
\end{array}
\right) ,\nonumber\\
\label{eq:mixingp}
\end{eqnarray}
where $\pi_p$ and $\pi_p^{\prime}$ in the left-hand side are the
physical states while $\pi$ and $\pi^\prime$ in the right-hand side
denote the $q\bar{q}$ states and $qq\bar{q}\bar{q}$ states,
respectively. In the present analysis we identify $\pi_p$ as
$\pi(140)$. As was shown in
Refs.~\cite{Fariborz:2005gm,Fariborz:2007ai}, the above pseudoscalar
mixing angle $\theta_\pi$ relates to the pion decay constant and
vacuum expectation values (VEVs) of the $q\bar{q}$ and
$qq\bar{q}\bar{q}$ scalar fields. In the chiral limit, we have 
\begin{eqnarray}
F_\pi \cos \theta_\pi & = & 2 v_2 \ ,\;\;\;\;\;\;
F_\pi \sin \theta_\pi = -2 v_4 \ ,
\label{eq:relationfpivev}
\end{eqnarray}
where $F_\pi = 130.41~$MeV~\cite{Nakamura:2010zzi} stands for the
decay constant of $\pi(140)$ and $v_2$ and $v_4$ denote the VEVs of
the $q\bar{q}$ and the $qq\bar{q}\bar{q}$ scalar fields,
respectively.

Using the method shown in Ref.~\cite{Fariborz:2005gm}, we obtain the
relation 
among the scalar meson-$\pi$-$\pi$ coupling constant $g_{f_j\pi\pi}$,
the mixing matrices and the scalar mass as: (for a derivation, see
Appendix~\ref{app:coupling})
\begin{eqnarray}
g_{f_j\pi\pi} & = & \frac{\sqrt{2}}{F_\pi}[\cos\theta_\pi
(U_f)_{ja} - \sin\theta_\pi(U_f)_{jc}]m_{f_j}^2 \, ,
\label{eq:couplingsigma2pi}
\end{eqnarray}
which 
is similar to the $\sigma$-$\pi$-$\pi$ coupling of
Ref.~\cite{Black:2000qq} except the mixing angle included and the
chiral limit taken. We also obtain the relation for the four-pion
coupling constant 
$g_{\pi\pi\pi\pi}$ as (see
Appendix \ref{app:coupling})
\begin{eqnarray}
g_{\pi\pi\pi\pi}
& = &
\dfrac{6}{ F_\pi^2 }
\sum_{j=1}^{4}
\left[
\cos\theta_\pi
(U_f)_{ja}
-
\sin\theta_\pi
(U_f)_{jc}
\right]^2
m_{f_j}^2
\ .\nonumber\\
\label{eq:rela_4}
\end{eqnarray}
Making use of the relations (\ref{eq:couplingsigma2pi}) 
and (\ref{eq:rela_4})
together with the orthonormal conditions,
\begin{eqnarray} 
&& \sum_{j=1}^4 (U_{ja})^2 = \sum_{j=1}^4 (U_{jc})^2 =1 \, , \nonumber\\
&& \sum_{j=1}^4 U_{ja} U_{jc} =0\, ,
\end{eqnarray}
we obtain the following sum rules:  
\begin{eqnarray}
\sum_{j=1}^4 \frac{g_{f_j\pi\pi}^2}{m_{f_j}^2} & = &
\frac{1}{3}g_{\pi\pi\pi\pi} \ , \label{eq:couplingsum1} \\
\sum_{j=1}^4 \frac{g_{f_j\pi\pi}^2}{(m_{f_j}^2)^2} & = &
\frac{2}{F_\pi^2}\ . \label{eq:couplingsum}
\end{eqnarray}
We should note that, as will be shown in the next
section, these sum rules guarantee that the $\pi$-$\pi$ scattering
amplitude satisfies the low-energy theorem of the Nambu-Goldstone (NG)
bosons.

\section{Sigma meson in the $\pi$-$\pi$ scattering}

\label{sec:scattering}

In this section, we determine the sigma meson mass $m_\sigma$ and
$\sigma$-$\pi$-$\pi$ coupling constant $g_{\sigma\pi\pi}$ using the
isospin zero ($I=0$), $S$-wave $\pi$-$\pi$ scattering data. 

The matrix element for the isospin zero $\pi$-$\pi$ scattering is
expressed as 
\begin{eqnarray}
A^{I=0} & = & 3 A(s,t,u) + A(t,u,s) +
A(u,s,t),\label{eq:decayscattering}
\end{eqnarray}
where $s, t$ and $u$ are the Mandelstam variables and the
amplitude $A(s,t,u)$ is defined in such a way that the invariant
amplitude for $\pi_i(p_1) + \pi_j(p_2) \rightarrow \pi_k(p_3)
+ \pi_l(p_4)$ is decomposed as 
\begin{eqnarray}
\delta_{ij}\delta_{kl} A(s,t,u) + \delta_{ik}\delta_{jl} A(t,s,u)
+ \delta_{il}\delta_{jk} A(u,t,s) \ .
\end{eqnarray}
The partial wave amplitude is obtained as 
\begin{eqnarray}
T^{I=0}_L & = & \frac{1}{2}\rho(s)\int_{-1}^{+1}d\cos\theta
P_L(\cos\theta)A^{I=0}(s,t,u),
%\label{partial}
\end{eqnarray}
where $\rho(s) = |\mathbf{q}(s)|/(16\pi\sqrt{s})$ with 
$|\mathbf{q}(s)| = \sqrt{s-4m_\pi^2} / 2$ and 
$P_L(\cos\theta)$ is the Legendre function.

In the present model, the scattering matrix $A(s,t,u)$ is expressed as
\begin{eqnarray}
A(s,t,u) & = & -\frac{1}{3}g_{\pi\pi\pi\pi} - \sum_{j=1}^4
g_{f_j\pi\pi}^2\frac{1}{s-m_{f_j}^2} \ .
\label{eq:full contribution}
\end{eqnarray}
It should be noticed that the sum rule~(\ref{eq:couplingsum}) implies
that 
\begin{equation}
A(s,t,u) = \frac{2s}{F_\pi^2} \ ,
\quad \mbox{for} \ s \ll m_\sigma^2 \ , \label{eq:letNG}
\end{equation}
consistently with the low-energy theorem of the NG bosons.
After the partial wave and the isospin projection, the matrix element
for the iso-singlet $S$-wave channel is obtained as
\begin{widetext}
%\onecolumn{
\begin{eqnarray}
T^{I=0}_{L=0}(s) & = & \rho(s)\left[-\frac{5}{3}g_{\pi\pi\pi\pi} +
2\frac{1}{s-4m_\pi^2}\sum_{j=1}^4 g_{f_j\pi\pi}^2\ln\left(\frac{s +
m_{f_j}^2 - 4m_\pi^2}{m_{f_j}^2}\right) - 3 \sum_{j=1}^4
g_{f_j\pi\pi}^2\frac{1}{s-m_{f_j}^2}\right]\ ,
\label{eq:partialscattering}
\end{eqnarray}
%}
\end{widetext}
%\twocolumn
where the second term comes from the $t$- and $u$-channel scalar
exchange contributions, and the third term comes from the $s$-channel
exchange. In this expression, $m_\pi$ stands for the $\pi(140)$ mass.

Since the rescattering effect should be properly included in the
energy region above $800$~MeV~\cite{Harada:1995dc}, the above
amplitude is applicable only in the low energy region.
Actually, in the calculation of the $D_1 \to D \pi \pi$ decay in
section \ref{sec:heavydecay}, the 
exchanged energy is $\sqrt{s} < 560$~MeV, so that, among the
exchanged four scalar mesons, the sigma meson gives a dominant
contribution. One might naively eliminate the contributions of
$f_{p2}$, $f_{p3}$ and $f_{p4}$ from the second and third terms in
Eq.~(\ref{eq:full contribution}) to reduce the number of parameters. 
However, such a truncated amplitude cannot reproduce the low-energy
theorem in Eq.~(\ref{eq:letNG}) obtained as a consequence of the chiral
symmetry. Then, instead of naively eliminating the scalar mesons other than sigma, we make an
expansion of the amplitude in terms of $s/m_{f_j}^2$ and
$m_\pi^2/m_{f_j}^2$ ($j=2,3,4$).
As a result, the partial wave amplitude $T_{L=0}^{I=0}(s)$ is reduced to
\begin{widetext}
\begin{eqnarray}
T_{L=0}^{I=0}(s)& = & \rho(s)\left[-\frac{5}{3}g_{\pi\pi\pi\pi} +
2\frac{1}{s-4m_\pi^2} g_{\sigma\pi\pi}^2\ln\left(\frac{s +
m_{\sigma}^2 -
4m_\pi^2}{m_{\sigma}^2}\right) - 
 3 g_{\sigma\pi\pi}^2\frac{1}{s-m_{\sigma}^2} \right. \nonumber\\
& & \left. \;\;\;\;\;\;\;\;\; + \sum_{j=2}^4
\left(5\frac{g_{f_j\pi\pi}^2}{m_{f_j}^2} -
\frac{g_{f_j\pi\pi}^2}{(m_{f_j}^2)^2}\left(s-4m_\pi^2\right) + 3
\frac{g_{f_j\pi\pi}^2}{(m_{f_j}^2)^2}s + \cdots\right) \right] \ ,
\end{eqnarray}
%\end{widetext}
where dots stand for the higher order terms.

Making use of Eqs.~(\ref{eq:couplingsum1}) and (\ref{eq:couplingsum}),
we arrive at 
%\begin{widetext}
\begin{eqnarray}
T_{L=0}^{I=0}(s) & = &
\rho(s)\left[-5\frac{g_{\sigma\pi\pi}^2}{m_{\sigma}^2} +
2\frac{1}{s-4m_\pi^2} g_{\sigma\pi\pi}^2\ln\left(\frac{s +
m_{\sigma}^2 -
4m_\pi^2}{m_{\sigma}^2}\right) - 3 g_{\sigma\pi\pi}^2\frac{1}{s-m_{\sigma}^2} +
\left(\frac{2}{F_\pi^2}-\frac{g_{\sigma\pi\pi}^2}{(m_{\sigma}^2)^2}\right)\left(2s+4m_\pi^2\right)\right] \,. \nonumber\\
\label{eq:pipiscatteringpartial}
\end{eqnarray}
\end{widetext}
We would like to stress that the amplitude
(\ref{eq:pipiscatteringpartial}) depends on only two undetermined
parameters $m_\sigma$ and $g_{\sigma\pi\pi}$, while the one in 
Eq.~(\ref{eq:partialscattering}) includes nine 
($m_{f_i}, g_{f_i\pi\pi}$ and $g_{\pi\pi\pi\pi}$). This low energy
reduction is achieved by using the relations in
Eqs.~(\ref{eq:couplingsum1}) and (\ref{eq:couplingsum}) which 
are the consequences of the chiral symmetry. Thus, for studying the $D_1 \rightarrow D \pi \pi$ decay rate,
it is enough to determine the parameters $m_\sigma$ and
$g_{\sigma\pi\pi}$.

To study the $\pi$-$\pi$ scattering we include the finite width effect
in the sigma meson propagator. Here, we use the modified Breit-Wigner
prescription where the width effect in the sigma meson propagator is
taken to be momentum dependent, that is, we make the substitution 
\begin{eqnarray}
\dfrac{1}{ s - m_\sigma^2 } \rightarrow \dfrac{1}{ s - m_\sigma^2 +
i m_\sigma \Gamma_\sigma(s)}.\label{eq:renormscalatprop}
\end{eqnarray}
The sigma meson width has several expressions in literature. 
In the present analysis, we take 
the width as the imaginary part of the
$\pi$-$\pi$ loop contribution to the $\sigma$ meson self-energy in the
linear sigma model
\begin{eqnarray}
\Gamma(s) = \frac{3 g_{\sigma\pi\pi}^2}{32\pi m_\sigma} 
\sqrt{1- \frac{4m_\pi^2}{s} } \ . \label{eq:widthsigma1}
\end{eqnarray}

\begin{figure}[htbp]
\begin{center}
\includegraphics[scale=0.8]{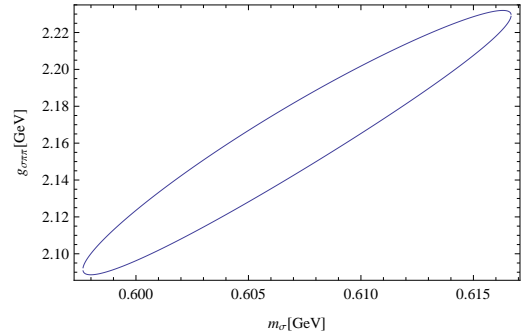}
%\;\;\;\;\;\;
%\includegraphics[scale=0.8]{contourS.eps}
\end{center}
\caption[]
{ Allowed region of $m_\sigma$ and $g_{\sigma\pi\pi}$ at $1\sigma$
% level. for Model I (left panel) and Model II (right panel).
} \label{fig:corre}
\end{figure}

Now, we fit the sigma meson mass $m_\sigma$ and the coupling constant
$g_{\sigma\pi\pi}$ from the $\pi$-$\pi$ scattering data given in
Refs.~\cite{Alekseeva:1982uy,Grayer:1974cr} in the low energy region below
$560$~MeV. The best fitted values are obtained 
as~\footnote{ When we use the parameterization of the width of the
  $\sigma$ propagator as $ \Gamma(s) = \frac{3
    g_{\sigma\pi\pi}^2}{32\pi s} \sqrt{1- \frac{4m_\pi^2}{s}
  }$~\cite{Black:2000qq}, 
  the best fitted values are $m_\sigma = 586 \pm 7 \, \mbox{MeV}$,
  $\vert g_{\sigma\pi\pi} \vert = 1.90 \pm 0.04 ~ \mbox{GeV}$ with 
  $\chi^2/{\rm d.o.f.} = 7.05/12 = 0.59$. On the other hand, when we use $ \Gamma = \frac{3 g_{\sigma\pi\pi}^2}{32\pi m_\sigma^2} \sqrt{1- \frac{4m_\pi^2}{m_\sigma^2} }$ we obtain $m_\sigma = 597 \pm 8 \, \mbox{MeV}$,
    $\vert g_{\sigma\pi\pi} \vert = 2.05 \pm 0.06 ~ \mbox{GeV}$ with 
    $\chi^2/{\rm d.o.f.} = 5.23/12 = 0.44$.
}
\begin{eqnarray}
& & m_\sigma = 606 \pm
9 \, \mbox{MeV}\, , \;\;\; 
\vert g_{\sigma\pi\pi} \vert = 2.16 \pm 0.07
~ \mbox{GeV}\, ,\nonumber\\
& & \frac{\chi^2}{\rm d.o.f.} = \frac{3.48}{12} =
0.29 \ . \label{eq:fit1}
\end{eqnarray}
Note that we can determine the absolute value of $g_{\sigma\pi\pi}$,
the sign of which becomes relevant in $D_1 \rightarrow D \pi\pi$
decay. Here we obtained the $g_{\sigma\pi\pi}$ in the chiral limit. We expect the correction from the inclusion of the pion mass is on the order of $m_\pi^2/m_\sigma^2$~\cite{Black:2000qq}, which is about 5\%.
%%%%%%%%%%%%%%%%%%%%%%%%%%%%%%%%%%%%%%%%%
\begin{figure}[htbp]
\begin{center}
\includegraphics[scale=0.8]{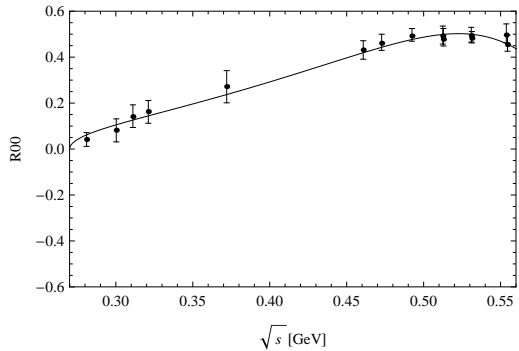}
\end{center}
\caption[]
{ Best fitted curve of the real part of the $I=0$ $S$-wave $\pi$-$\pi$
scattering  amplitude. The Data are taken from
Refs.~\cite{Alekseeva:1982uy,Grayer:1974cr}. } \label{fig:pipi} 
\end{figure}
%%%%%%%%%%%%%%%%%%%%%%%%%%%%%%%%%%%%%%%%%%
We show the best fit curve 
in Fig.~\ref{fig:pipi} with the allowed region of $m_\sigma$ and
$g_{\sigma\pi\pi}$ at $1\sigma$ level shown in Fig.~\ref{fig:corre}. 
We conclude that our model can reproduce the $\pi$-$\pi$ scattering
data below $560$~MeV quite well. 
%%%%%%%%%%%%%%%%%%%%%%%%%%%%%%%%%%%%%%%%%%
%%%%%%%%%%%%%%%%%%%%%%%%%%%%%%%%%%%%%%%%%%

\section{Effective Lagrangian for heavy-light mesons with chiral
partner structure} 

\label{sec:heavyL}

In this section, we introduce an effective Lagrangian for the
heavy-light mesons coupling to the light mesons based on the heavy
quark symmetry 
combined with the chiral symmetry. For constructing the Lagrangian
invariant under the linearly realized chiral symmetry, we need to
include the chiral partner to the lowest lying heavy quark multiplet
of $H=(D^\ast, D)$. In the present analysis, we regard the doublet
$G=(D_1, D_0^{\ast})$ as the chiral partner to the $H$
doublet~\cite{Nowak:1992um,Bardeen:1993ae}. For constructing the
effective Lagrangian including these fields, it is essential to
consider the symmetry properties, especially the $U(1)_A$ charge, of
these heavy-light mesons. Since the mass spectra of the 
states in the $H$ and $G$ doublets are consistent with the theoretical
predictions based on the $c\bar{q}$ interpretation, it is
reasonable to regard them as the $c\bar{q}$ states.
Following 
Refs.~\cite{Nowak:1992um,Bardeen:1993ae,Bardeen:2003kt,Harada:2003kt},
we include the $H$ and $G$ doublets into the Lagrangian through 
\begin{eqnarray}
\mathcal{H}_{L} & = & \frac{1}{\sqrt{2}}[G + iH\gamma_5], \;\;\;\;\;
\mathcal{H}_{R} = \frac{1}{\sqrt{2}}[G - iH\gamma_5]
\ .
\label{eq:heavychiral}
\end{eqnarray}
In terms 
of the physical states, the $H$ and 
$G$ doublets are expressed as 
\begin{eqnarray}
H & = & \frac{1+v\hspace{-0.2cm}\slash}{2}\left[D^{\ast \,
\mu}\gamma_\mu + i D \gamma_5 \right], \nonumber\\
G & = &
\frac{1+v\hspace{-0.2cm}\slash}{2}\left[ - i D^{\mu}_{1}
\gamma_\mu\gamma_5 + D_0^{\ast} \right],\label{eq:hdphys}
\end{eqnarray}
with $v^\mu$ being the velocity of the heavy meson.

Under chiral transformation, the $\mathcal{H}_L$ and $\mathcal{H}_R$
fields transform as  
\begin{eqnarray}
\mathcal{H}_L & \to & \mathcal{H}_L g_L^\dag, \;\;\;\;\;
\mathcal{H}_R \to \mathcal{H}_R g_R^\dag.
\end{eqnarray}
under the axial $U(1)_{A}$ symmetry, $\mathcal{H}_L$ and
$\mathcal{H}_R$ transform as 
\begin{eqnarray}
\mathcal{H}_L \rightarrow \mathcal{H}_L e^{-i\alpha}, \;\;\;\;\;
\mathcal{H}_R \rightarrow \mathcal{H}_R e^{i\alpha}.
\end{eqnarray}
Using the chiral and $U(1)_{A}$ transformation properties of
heavy-light meson fields and light meson fields, we construct an
effective Lagrangian describing the interactions among the 
heavy-light mesons and the light mesons. In the present construction,
we only include the minimal number of terms which are responsible for
our following analysis of the $D_1 \to D \pi\pi$ decay. Then the
Lagrangian is written as  
\begin{widetext}
\begin{eqnarray}
\mathcal{L}_{\rm heavy} & = & \frac{1}{2} {\rm Tr}\left[
\bar{\mathcal{H}}_L i(v \cdot \partial) \mathcal{H}_L\right] +
\frac{1}{2}{\rm Tr} \left[\bar{\mathcal{H}}_R i(v\cdot
\partial)\mathcal{H}_R\right]  -\frac{\Delta}{2} {\rm Tr}\left[\bar{\mathcal{H}}_L \mathcal{H}_L
+ \bar{\mathcal{H}_R} \mathcal{H}_R\right] \nonumber\\
& & {} -\frac{g_\pi}{4}{\rm Tr}\left[ M^\dagger
\bar{\mathcal{H}}_L\mathcal{H}_R + M
\bar{\mathcal{H}}_R\mathcal{H}_L\right] + i\frac{g_A}{2F_\pi} {\rm
  Tr}\left[ \gamma^5\partial\hspace{-0.2cm}\slash M^\dagger
  \bar{\mathcal{H}}_L \mathcal{H}_R -
  \gamma^5\partial\hspace{-0.2cm}\slash M \bar{\mathcal{H}}_R
  \mathcal{H}_L \right] , 
\label{eq:heavylagranchiral}
\end{eqnarray}
where $\Delta, g_\pi$ and $g_A$ are parameters.
In terms of the $H$ and $G$ doublets, the effective Lagrangian
(\ref{eq:heavylagranchiral}) is rewritten as 
%\begin{widetext}
\begin{eqnarray}
{\cal L}_{\rm heavy} & = & \frac{1}{2}{\rm Tr}\left[ - \bar{H}iv\cdot
  \partial H + \bar{G}iv\cdot \partial G \right] -
\frac{\Delta}{2}{\rm Tr}\left[ - \bar{H} H + \bar{G} G \right]
\nonumber\\ 
& & - \frac{g_\pi}{8}{\rm Tr}\left[\left(M +
  M^\dag\right)\left(\bar{G} G + \bar{H} H\right) - i\left(M -
  M^\dag\right)\left(\bar{H} G - \bar{G} H\right)\gamma^5\right]
\nonumber\\ 
& & + i\frac{g_A}{4 F_\pi}{\rm Tr}\left[-\left(\partial
\hspace{-0.2cm}\slash M - \partial \hspace{-0.2cm}\slash
M^\dag\right)\left(\bar{G} G + \bar{H} H\right)\gamma^5 -
i\left(\partial \hspace{-0.2cm}\slash M + \partial 
\hspace{-0.2cm}\slash M^\dag\right)\left(\bar{H} G + \bar{G}
H\right)\right]. 
\label{eq:heavylagranphys}
\end{eqnarray}
\end{widetext}

\section{The sigma meson structure from $D_1 \to D \pi\pi$ decay}

\label{sec:heavydecay}

In this section, we study the quark contents of the $\sigma$ meson
through the $D_1 \to D \pi\pi$ decay.

\subsection{Determination of the parameters in the heavy meson
  effective Lagrangian}

In this subsection, we determine the parameters in the Lagrangian
(\ref{eq:heavylagranphys}) that are necessary for the numerical
calculation of the $D_1 \to D \pi\pi$ decay width.

First, since the kinetic terms of $H$ and $G$ doublets have the
opposite signs, the $\Delta$ term in Eq.~(\ref{eq:heavylagranphys})
shifts the masses of $H$ doublet and $G$ doublet to the same
direction. Then, in the following analysis, we take $\Delta = 0$
without loss of generality.

Second, we fix the value of the combination $g_A \cos\theta_\pi$ from
the partial width 
for $D^{\ast } \to D \pi$ decay. From the Lagrangian
(\ref{eq:heavylagranphys}), the $D^{\ast}$-$D$-$\pi$ interaction is
extracted as 
\begin{eqnarray}
{\cal L} _{D^{\ast} D \pi} & = & - \frac{ig_A}{ F_\pi} D \partial_\mu \Phi D^{\ast \mu \dagger} + {\rm H.c.}.
\label{eq:LagDastDpi}
\end{eqnarray}
{}From this the $D^{\ast} \to D \pi$ is expressed as
\begin{eqnarray}
\Gamma(D^{\ast \, +} \rightarrow D \pi) & = & \frac{(g_A
\cos\theta_\pi)^2}{4\pi F_\pi^2}\frac{m_H^2}{m_{D^{\ast}}^2}
\left|{\bf p_\pi}\right|^3
\ ,
\label{eq:widthdst}
\end{eqnarray}
where $\left|{\bf p_\pi}\right|$ is three-momentum of outgoing 
pion and $m_{H}$ is the average mass defined as
$m_{H} = ( 3 m_{D^\ast} + m_{D} )/4$.
Using the central value for 
$\Gamma_{D^{\ast + }} = 96~$KeV~\cite{Nakamura:2010zzi} we get 
\begin{equation}
\left\vert g_A\cos\theta_\pi \right\vert= 0.56 
\ .
\end{equation}

Third, we determine the combination $g_\pi\cos\theta_\pi$ using 
the $D_0^{\ast} \to D \pi$ decay. The relevant terms in the 
Lagrangian (\ref{eq:heavylagranphys}) are given by
\begin{eqnarray}
{\cal L}_{D_0^{\ast}D\pi} & = & \frac{ig_\pi}{2}
D \Phi D_{0}^{\ast \, \dagger} + {\rm H.c.}, \label{eq:intd0stdpi}
\end{eqnarray}
From this Lagrangian, the width for $D_0^{\ast} \to D \pi$ decay is
expressed as 
\begin{eqnarray}
\Gamma(D_0^{\ast} \to D \pi) & = & \frac{3}{2}
\frac{m_H m_G}{2 \pi m_{D_0^{\ast}}^2}
\frac{(g_\pi\cos\theta_\pi)^2}{4} 
\left|\mathbf{p}_\pi\right| 
\ ,
\label{eq:d0dspidecay}
\end{eqnarray}
where $m_{G}$ is the average mass defined as
$m_{G} = ( 3 m_{D_1} + m_{D_0^\ast} )/4$.
Using the central value of the data 
$\Gamma(D_0^{\ast}) = 267$~MeV~\cite{Nakamura:2010zzi} 
and assuming 
that $D_0^{\ast}$ decays dominantly to $D \pi$, 
we obtain 
\begin{equation}
\left\vert g_\pi \cos \theta_\pi \right\vert = 3.61
\ .
\end{equation}

With the numerical value of $\vert g_\pi \cos\theta_\pi \vert$ estimated above we predict the $D_1 \to D^{\ast} \pi$ decay width as a check
of the validity of the heavy quark expansion applied here. From the
Lagrangian (\ref{eq:heavylagranphys}) we obtain the relevant
Lagrangian as 
\begin{eqnarray}
{\cal L}_{D_1D^{\ast}\pi} & = & - \frac{ig_\pi}{2}
D^{\ast}_\mu \Phi D_1^{ \mu \dagger} + {\rm H.c.}, \label{eq:intd1dstpi}
\end{eqnarray}
which yields the following expression of the $D_1 \to D^{\ast} \pi$
decay width:
\begin{eqnarray}
\Gamma(D_1 \to D^{\ast} \pi) & = & 
\frac{3}{2}\frac{m_H m_G}{2 \pi m_{D_1}^2}
\frac{(g_\pi\cos\theta_\pi)^2}{4}\left|\mathbf{p}_\pi\right| \, .
\label{eq:d1dspidecay}
\end{eqnarray}
Using the central values for the relevant particle masses we obtain
\begin{eqnarray}
\Gamma(D_1 \to D^{\ast} \pi) & = & 224.0~{\rm MeV}.
\end{eqnarray}
Comparing this with the data 
$\Gamma(D_1) = 383^{+107}_{-75}\pm 74$~\cite{Nakamura:2010zzi} 
we conclude that our prediction, based on the heavy quark limit, is
consistent with the data. This implies that the heavy quark limit
taken here is a reasonable approach for the $H$ and $G$ doublets.

\subsection{$D_1 \to D \pi\pi$ decay}

In this subsection, using the model explained so far, we study the
$D_1 \rightarrow D \pi \pi$ decay. We show the relevant diagrams in
the present model in Fig.~{\ref{fig:feyndiag}}. 
\begin{widetext}
\begin{center}
\begin{figure}[htbp]
%\begin{center}
\includegraphics[scale=0.6]{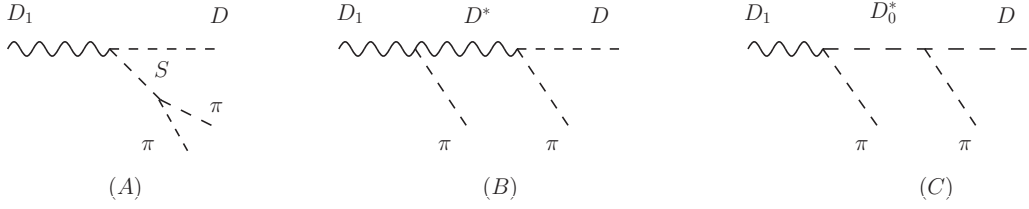}
%\end{center}
\caption[Feynman diagrams contributing to $D_1 \to D \pi\pi$ decay in
  our model.]{% 
Feynman diagrams contributing to $D_1 \to D \pi\pi$ decay in our
model.} \label{fig:feyndiag} 
\end{figure}
\end{center}
The matrix element for this process is
straightforwardly written as 
\begin{eqnarray}
{\cal M} & = & - \sqrt{2 m_{G}m_{H}} \frac{g_A}{F_\pi}
\epsilon_\mu(v) \left\{\sum_{i = 1}^{4} g_{f_i\pi\pi} \frac{(p_{\pi_1} +
p_{\pi_2})^\mu}{s - m_{f_i}^2 + im_{f_i}\Gamma_{f_i}(s)} (U_f^{-1})_{ai} \right. \nonumber
\\
& &
\left. \;\;\;\;\;\;\;\;\;\;\;\;\;\;\;\;\;\;\;\;\;\;\;\;\;\;\;\;\;\;\;\;\;\;\;\;\;\;
     {} + \frac{g_\pi}{2\sqrt{2}} \cos^2\theta_\pi 
\frac{p^\mu_{\pi_2} - v^\mu v\cdot p_{\pi_2}}{v\cdot k_{D^{\ast}} +
  i\Gamma_{D^*}/2} + \frac{g_\pi}{2\sqrt{2}} \cos^2\theta_\pi 
\frac{p^\mu_{\pi_1}}{v\cdot k_{D_0^{\ast}} + i\Gamma_{D_0^*}/2}
\right\} , \label{eq:decaymatrix} 
\end{eqnarray}
%\end{widetext}
where $s = (p_{\pi_1} + p_{\pi_2})^2 = m_{\pi\pi}^2$ with $p_{\pi_1}$
and $p_{\pi_2}$ as the momenta of pions in the final states,
$\epsilon_\mu(v)$ is the polarization vector of $D_1$ meson satisfying
$\sum_{\rm polarization} \epsilon_\mu \epsilon^{\ast}_\nu = 
-(g_{\mu\nu}-v_\mu v_\nu)$, and the residual momenta $k_{D^{\ast}}$ and $k_{D_0^{\ast}}$ are defined
by $p_{D^*}^\mu = m_{D^\ast}v^\mu + k_{D^\ast}^\mu$ and $p_{D_0^{\ast}}^\mu = m_{D_0^{\ast}}v^\mu + k_{D_0^{\ast}}^\mu$, respectively.

We note that, since the maximum value of
the $\pi\pi$ invariant mass is restricted as $m_{\pi\pi} \le m_{D_1} -
m_D \simeq 560$~MeV, the contribution from the lightest scalar meson,
the sigma meson is dominant. Along the same method as that was used in
the $\pi$-$\pi$ scattering 
case, using the sum rules (\ref{eq:couplingsum1}) and
(\ref{eq:couplingsum}), the matrix element (\ref{eq:decaymatrix}) is
reduced as (for a 
derivation, see Appendix \ref{app:reduction})
%\begin{widetext}
\begin{eqnarray}
{\cal M} & = & - \sqrt{2m_{G}m_{H}} \frac{g_A\cos\theta_\pi}{F_\pi}
\epsilon_\mu(v) \left\{ h \, g_{\sigma\pi\pi} \frac{(p_{\pi_1} +
  p_{\pi_2})^\mu}{s - m_{\sigma}^2 + im_{\sigma}\Gamma_{\sigma}(s)} 
- (p_{\pi_1} + p_{\pi_2})^\mu\left(\frac{\sqrt{2}}{F_\pi} - 
\frac{h\, g_{\sigma\pi\pi}}{m_{\sigma}^2}
\right) \right. \nonumber\\
& & \left.\qquad\qquad\qquad\qquad\qquad\qquad {} +
\frac{g_\pi\cos\theta_\pi}{2\sqrt{2}} \frac{p^\mu_{\pi_2} - v^\mu
  v\cdot p_{\pi_2}}{v\cdot k_{D^{\ast}} + i\Gamma_{D^*}/2} + \frac{g_\pi
  \cos\theta_\pi}{2\sqrt{2}}  
\frac{p^\mu_{\pi_1}}{v\cdot k_{D_0^{\ast}} + i\Gamma_{D_0^*}/2}  \right\} \, ,
\label{eq:decaymatrixsum}
\end{eqnarray}
%\end{widetext}
where the mixing parameter $h$ is defined as
\begin{eqnarray}
h & = & \frac{(U_f^{-1})_{a1}}{\cos\theta_\pi}. \label{eq:defh}
\end{eqnarray}
In the amplitude (\ref{eq:decaymatrixsum}), the values of $\vert g_\pi
\cos \theta_\pi \vert$ 
and $\vert g_A \cos\theta_\pi \vert$ 
as well as those of $m_\sigma$ and $\vert g_{\sigma\pi\pi} \vert$
are phenomenologically determined above. Then the amplitude
(\ref{eq:decaymatrixsum}) depends only on the mixing parameter
$h$. When the physical pion is a pure $q\bar{q}$ state, i.e.,
$\cos\theta_\pi = 1$, then $h=1$ implies that $\sigma$ meson is the
$q\bar{q}$ state while $h=0$ the pure $qq\bar{q}\bar{q}$ state.

Next, we make the isospin and the partial wave projection with respect
to the final 
two pions and pick up the $I = 0, S$-wave amplitude. This procedure
eliminates the contributions from $L = 2, 4, \cdots$ in diagrams (B)
and (C) in Fig.~\ref{fig:feyndiag}. \footnote{ 
We calculated the $D$-wave decay width which is less than 5\% of the
$S$-wave width. } The isospin projection gives the isospin singlet as 
\begin{eqnarray}
| \pi \pi, I =0 \rangle & = & \frac{1}{\sqrt{3}} \left| \pi^+
(p_1)\pi^-(p_2) + \pi^-(p_1) \pi^+(p_2) + \pi^0(p_1)\pi^0(p_2)
\right\rangle. \label{eq:isodecomp} 
\end{eqnarray}
And, the partial wave projection decompose the $D_1 \rightarrow D \pi
\pi$ decay matrix as 
\begin{eqnarray}
\frac{d \Gamma(D_1 \rightarrow D \pi \pi)}{dm_{\pi\pi}} &
= & \sum_L \frac{d \Gamma_L(D_1 \rightarrow D \pi \pi)}{dm_{\pi\pi}},
\end{eqnarray}
where
\begin{eqnarray}
\frac{d \Gamma_L(D_1 \rightarrow D \pi \pi)}{dm_{\pi\pi}}
& = & 2(2L+1)\frac{1}{2!}\frac{1}{64\pi^3
  m_{D_1}}\left(\frac{1}{2}\right)\frac{|\mathbf{p}_{\pi}|^2}{E_D}\sum_{\rm
  polarization} \left|{\cal
  M}_{L}(m_{\pi\pi}^2)\right|^2, \label{eq:difwidthdl} 
\end{eqnarray}
with $|\mathbf{p}_{\pi}| = \sqrt{m_{\pi\pi}^2-4m_\pi^2}/2$ as the
three momentum of the outgoing pions in the center of mass frame of two pions. ${\cal M}_{L}(m_{\pi\pi}^2)$
stands for the $D_1 \rightarrow D \pi \pi$ decay matrix element with the final two pions in $L$-wave, explicitly, 
\begin{eqnarray}
{\cal M}_{L}(m_{\pi\pi}^2) & = & \frac{1}{2}\int_{-1}^{1} d\cos
\theta P_L(\cos \theta){\cal
M}(m_{\pi\pi},\cos\theta),
\end{eqnarray}
with ${\cal M}$ given by Eq.~(\ref{eq:decaymatrixsum}) and $\theta$
being the angle between $\pi$ and $D$. In the following, we only
consider the $I = 0, S$-wave 
$d \Gamma( D_1 \rightarrow D(\pi\pi)_{I = 0, L = 0})/dm_{\pi\pi}$ 
to study the composition of the $\sigma$ meson. Taking into account of the sign ambiguities of
$g_{\sigma\pi\pi}$ and $g_\pi \cos\theta_\pi$, we show, in Fig.~\ref{fig:D1decayresult}, the resultant width
with $h = 0$ (solid line) and $h = 1$ (dotted line)
for four cases:
(a) $g_{\sigma\pi\pi} > 0$ and $g_\pi \cos\theta_\pi > 0 $;
(b) $g_{\sigma\pi\pi} > 0$ and $g_\pi \cos\theta_\pi < 0 $;
(c) $g_{\sigma\pi\pi} < 0$ and $g_\pi \cos\theta_\pi > 0 $;
and 
(d) $g_{\sigma\pi\pi} < 0$ and $g_\pi \cos\theta_\pi < 0 $. (
 Since $g_A \cos \theta_\pi$ is the overall factor of the amplitude,
 the sign ambiguity is irrelevant to the decay width.)
 This shows that, for $g_{\sigma\pi\pi} > 0$,
 the width for $h = 1$ is much larger than that for $h=0$. On the other hand, for $g_{\sigma\pi\pi} < 0$, the peak position is located around $450$~MeV for $h = 0$ while above $500$~MeV for $h = 1$. This suggests that, when the experimental data become available in the future, the mixing parameter $h$ together with the signs of $g_{\sigma\pi\pi}$ and $g_{\sigma\pi\pi}$ could be fitted.
 
%\begin{widetext}
\begin{center}
\begin{figure}
\includegraphics[width=7.0cm]{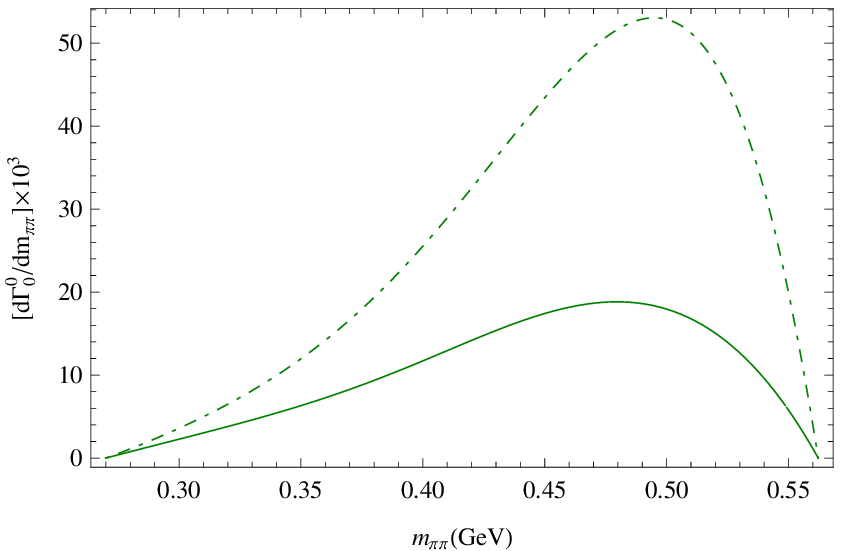}
\hspace{0.5cm}
\includegraphics[width=7.0cm]{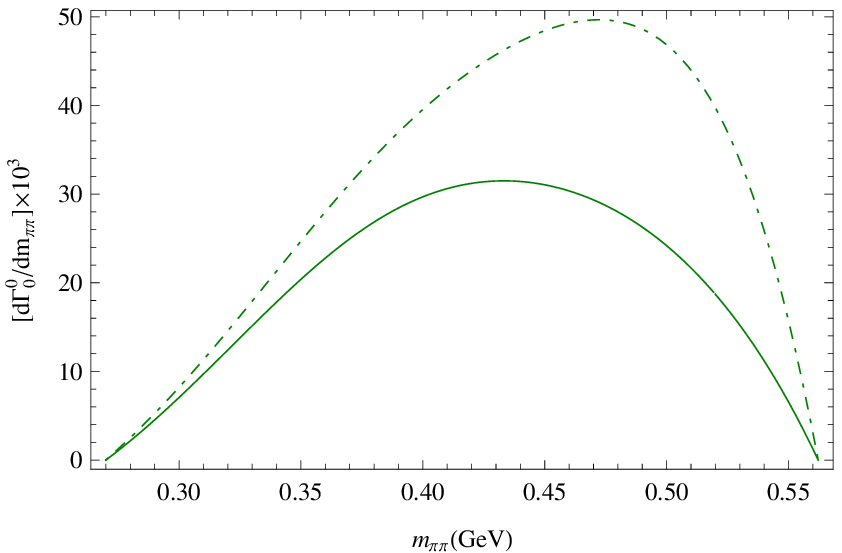}
\\
\includegraphics[width=7.0cm]{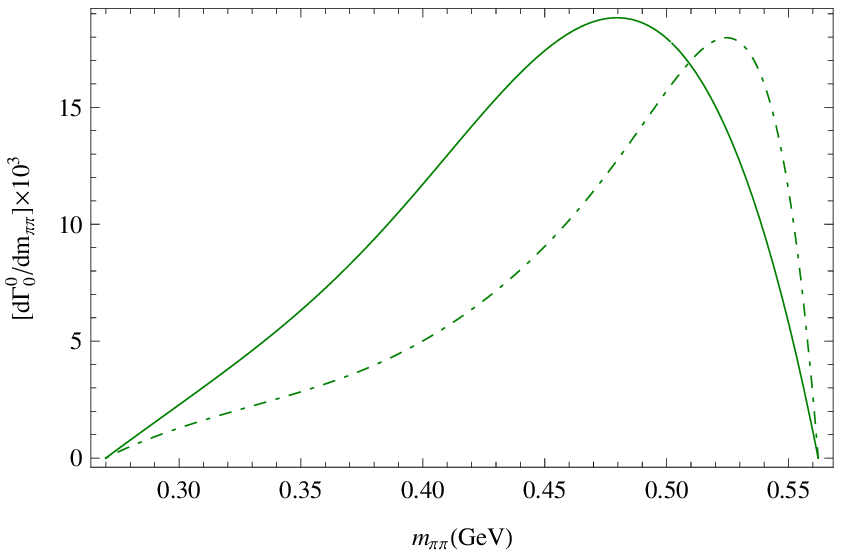}
\hspace{0.5cm}
\includegraphics[width=7.0cm]{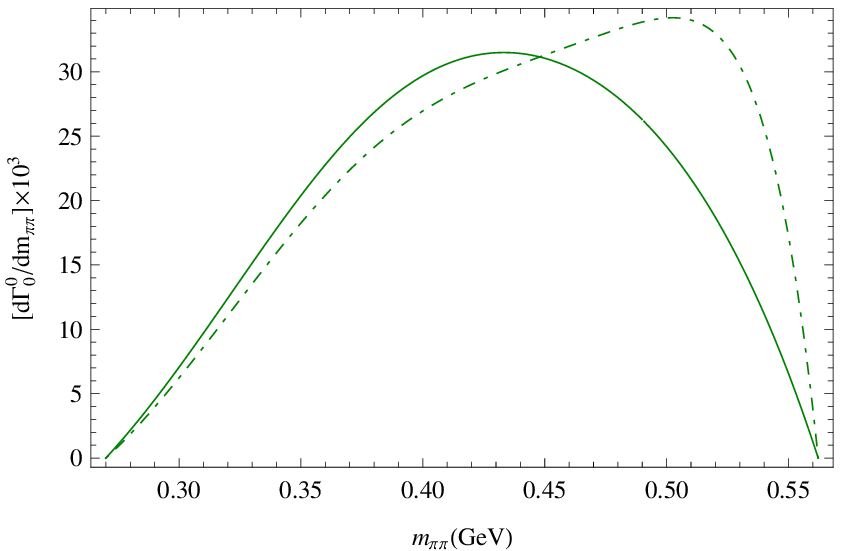}
\caption{$d\Gamma(D_1 \to D (\pi\pi)_{I=0,L = 0}) / dm_{\pi\pi} $ vs $m_{\pi\pi}$ with 
$h = 0$ (solid line) and
$h = 1$ (dotted line)
for $g_{\sigma\pi\pi} > 0$ and $g_\pi \cos\theta_\pi > 0$ 
(upper left panel),
$g_{\sigma\pi\pi} > 0$ and $g_\pi \cos\theta_\pi < 0$ 
(upper right panel),
$g_{\sigma\pi\pi} < 0$ and $g_\pi \cos\theta_\pi > 0$ 
(lower left panel),
$g_{\sigma\pi\pi} < 0$ and $g_\pi \cos\theta_\pi < 0$ 
(lower right panel).
}
\label{fig:D1decayresult}
\end{figure}
\end{center}
%\end{widetext} 
\vspace{2cm}
\end{widetext}

\section{A summary and discussions}

\label{sec:con}

We study how the composition of the lightest scalar meson $\sigma$
affects the $D_1 \rightarrow D \pi \pi $ decay.
We construct a linear sigma model for the $q \bar{q}$ 
and $qq \bar{q}\bar{q}$ chiral nonet fields, distinguishing two
nonets by their $U(1)_A$ charge.
We write down the effective interaction terms among
$c \bar{q}$-type heavy mesons ($D$, $D^\ast$, $D_0^\ast(2400)$,
$D_1(2430)$) and the scalar nonets, in which
the existence of the $U(1)_A$ symmetry implies that
only the $q\bar{q}$ component of the scalar meson couples to
the heavy mesons.
%-------------
%\vspace{0.4cm}
%\hrule width 3cm
%-------------

In the light quark sector, we use the most general form 
of the potential for the two chiral nonet fields.
Using the sum rules among the coupling constants and masses
of scalar mesons, we express the $\pi\pi$ scattering amplitude
in the low-energy region in terms of only two parameters, $m_\sigma$
and $g_{\sigma\pi\pi}$. We fit the values of $m_\sigma$ and
$g_{\sigma\pi\pi}$ 
to the $\pi\pi$ scattering data below 560\,MeV, which is the maximum
energy transferred to two pions in the $D_1 \rightarrow D \pi \pi $
decay. 
We fix the parameters in the effective interaction terms to heavy 
meson phenomenologically. We show how the $I = 0, S$-wave differential
decay rate,  
$d \Gamma( D_1 \rightarrow D (\pi\pi)_{I=0,L=0}) / d m_{\pi\pi}$,
depends on the mixing structure of the $\sigma$ meson: 
For $g_{\sigma\pi\pi} > 0$,
the width for $h = 1$ is much larger than that for $h=0$,
and the peak position moves to the higher energy region
for $g_{\sigma\pi\pi} < 0$.

In the theoretical calculation, we can include a part of the final
state interaction effect. The final state interaction among all the
three final particles and that between one pion and $D$ meson are
suppressed by the heavy meson mass therefore their effects are
small. For the final state interaction between the two pions, in the
$I=0, S$-wave channel,  the contribution from the bubble diagrams is
illustrated as  
\begin{widetext}
\begin{eqnarray}
\begin{array}{l}
    \hbox{\includegraphics[scale=0.45]{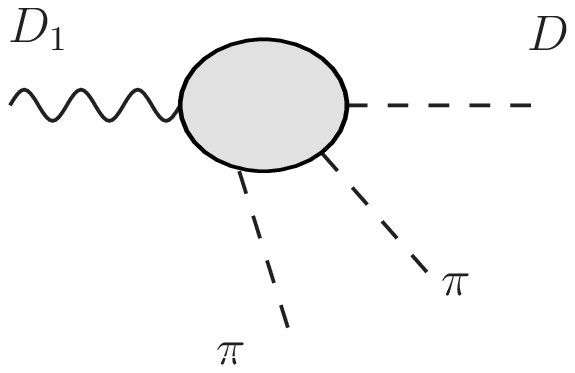}}
  \end{array} & = & \left\{
  \begin{array}{l}
    \hbox{\includegraphics[scale=0.4]{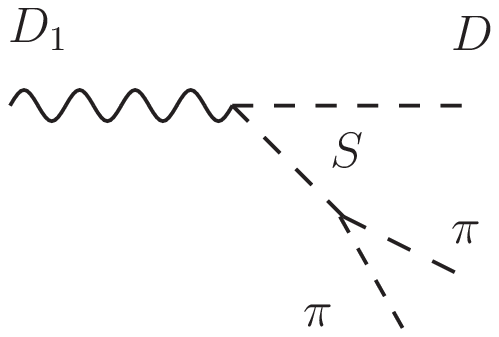}}
  \end{array} + \begin{array}{l}
      \hbox{\includegraphics[scale=0.4]{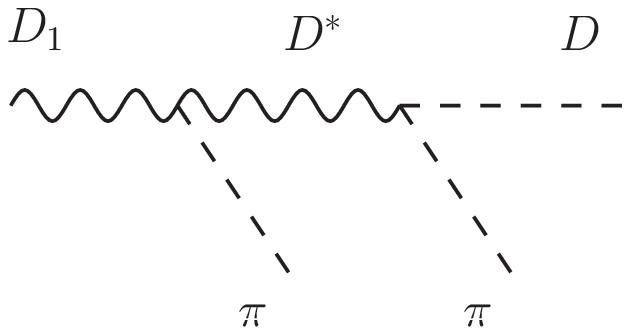}}
    \end{array} + \begin{array}{l}
        \hbox{\includegraphics[scale=0.4]{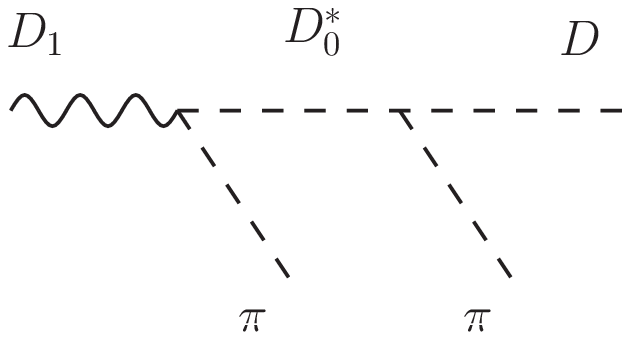}}
      \end{array}
\right\} \nonumber\\
& & \times \left\{
  \mathbf{1}
+ \begin{array}{l}
    \hbox{\includegraphics[scale=0.4]{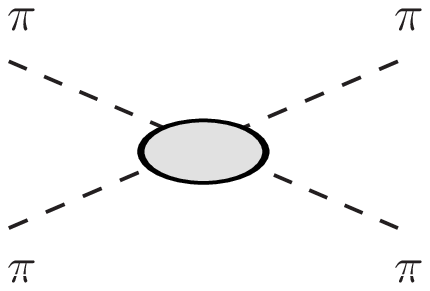}}
\end{array}
+ \begin{array}{l}
    \hbox{\includegraphics[scale=0.4]{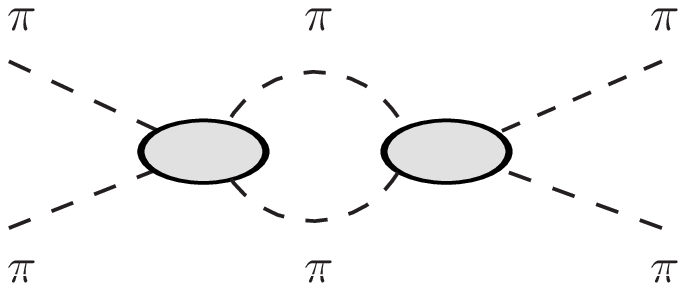}}
  \end{array}
+ \cdots \right\}\nonumber
\end{eqnarray}
\end{widetext}
The summation of the terms in the second parenthesis in the right hand
side of the above equation gives the $I = 0, S$-wave $S$-matrix for the $\pi$-$\pi$
scattering, therefore it just contributes a phase factor to the
$D_1 \to D \pi \pi$ decay matrix element and further, does not change
the partial width. In this sense, the correction to our results from
the final state interaction is small.

In the present analysis, the interaction terms among the heavy-light
mesons and light mesons is constructed in the heavy quark limit with
the linear 
realization of the chiral symmetry, therefore the $D_1 D \sigma$ and
$D^{\ast} D \pi$ 
coupling constants are related by the chiral symmetry.
Since the chiral symmetry is dynamically broken, there exists a
difference 
between the $D_1 D \sigma$ and $D^{\ast} D \pi$ coupling constants.
To show this chiral symmetry breaking effect, we typically rescale the
$D_1 D \sigma$ coupling constant by a factor two and illustrate out
results in Fig.~\ref{fig:D1decayresultrescale}. This shows that the
tendency of the differential width $d\Gamma(D_1 \to D(\pi\pi)_{I = 0,
  L = 0}) / dm_{\pi\pi} $ does not change. 

\begin{widetext}
\begin{center}
\begin{figure}
\includegraphics[width=7.5cm]{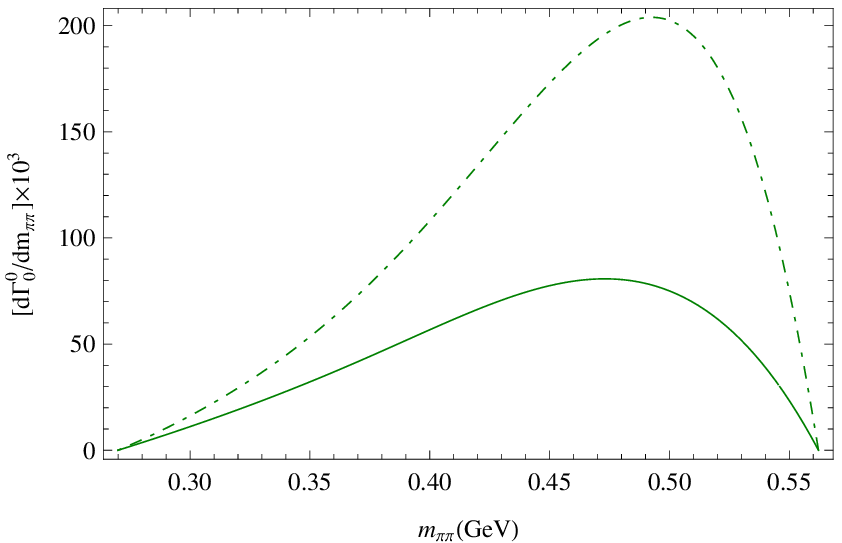}
\hspace{0.5cm}
\includegraphics[width=7.5cm]{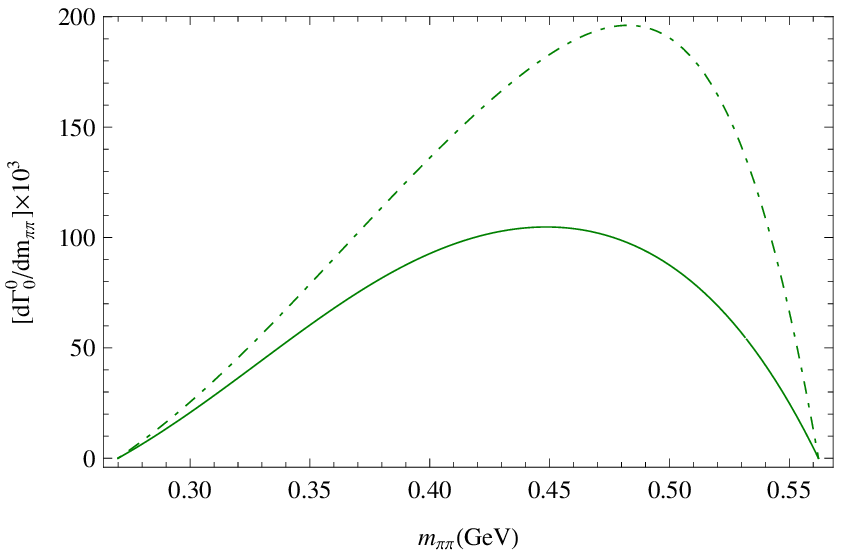}
\\
\includegraphics[width=7.5cm]{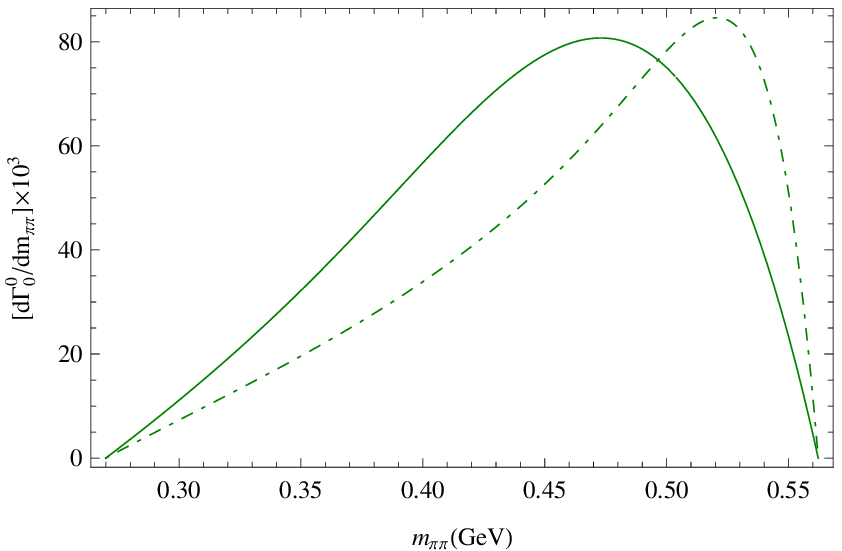}
\hspace{0.5cm}
\includegraphics[width=7.5cm]{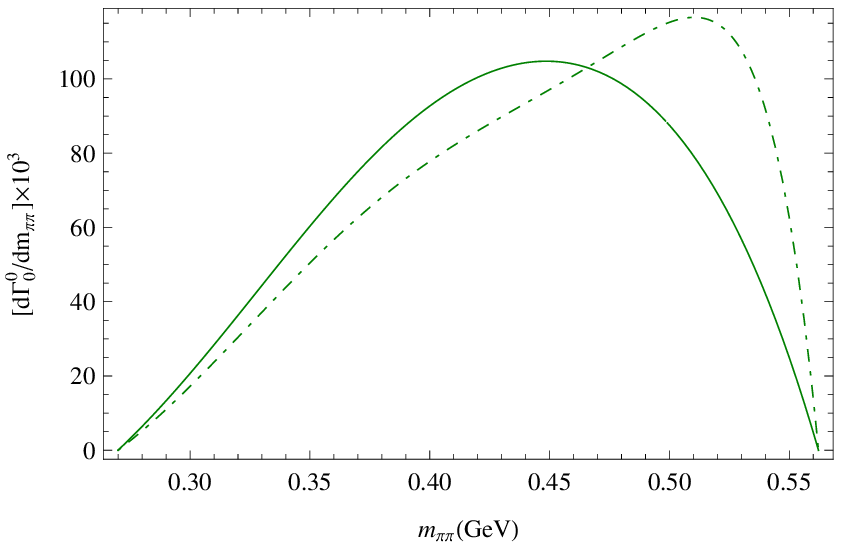}
\caption{
$d\Gamma(D_1 \to D (\pi\pi)_{I=0,L = 0}) / dm_{\pi\pi} $ vs $m_{\pi\pi}$ with 
$h = 0$ (solid line) and
$h = 1$ (dotted line)
for $g_{\sigma\pi\pi} > 0$ and $g_\pi \cos\theta_\pi > 0$ 
(upper left panel),
$g_{\sigma\pi\pi} > 0$ and $g_\pi \cos\theta_\pi < 0$ 
(upper right panel),
$g_{\sigma\pi\pi} < 0$ and $g_\pi \cos\theta_\pi > 0$ 
(lower left panel),
$g_{\sigma\pi\pi} < 0$ and $g_\pi \cos\theta_\pi < 0$ 
(lower right panel).
}\label{fig:D1decayresultrescale}
\end{figure}
\end{center}
%\end{widetext}

%\begin{widetext}
%%%%%%%%%%%%%%%%%%%%%%%%%%%%%%%%%%%%%%%%%%%%%%%%%%%%%%%%%%%%%%%%%%%%%%%
\acknowledgments

%\section*{Acknowledgments}

\label{ACK}

We would like to thank Prof. Daisuke Jido for his valuable
comments. This work is supported in part by Grant-in-Aid for
Scientific Research on Innovative Areas (No. 2104) ``Quest on New
Hadrons with 
Variety of Flavors'' from MEXT. The work of M.H. is supported in
part by the Grant-in-Aid for Nagoya University Global COE Program
``Quest for Fundamental Principles in the Universe: from Particles
to the Solar System and the Cosmos'' from MEXT, the JSPS
Grant-in-Aid for Scientific Research (S) $\sharp$ 22224003, (c)
$\sharp$ 24540266. The work of Y.M. is supported in part by the
National Science Foundation of China (NNSFC) under grant No.
10905060.

\newpage

%%%%%%%%%%%%%%%%%%%%%%%%%%%%%%%%%%%%%%%%%%%%%%%%%%%%%%%%%%%%%%%%%%%
\appendix

\section{Derivation of the coupling constants}

\label{app:coupling}

In this appendix, we derive the relations
(\ref{eq:couplingsigma2pi}) and (\ref{eq:couplingsum}) 
from the general
Lagrangian (\ref{eq:lagrlight}). Following
Ref.~\cite{Fariborz:2007ai}, we write the infinitesimal
transformation matrices for vector and axial-vector transformations
as $E_V$ and $E_A$, respectively. These infinitesimal matrices
satisfy the following relations:
\begin{eqnarray}
E_V^\dag & = &- E_V, \;\;\;\;\ E_A ^\dag = - E_A, \;\;\;\;\
{\rm Tr }{E_A} = 3i\alpha.\nonumber
\end{eqnarray}
Then, under the vector and axial-vector transformations, the relevant
meson fields transform as
\begin{eqnarray}
\delta _V \Phi & = & [E_V,\Phi ] ,\;\;\;\;\;\;\;\;\;\;\;\; \delta _V S =
[E_V,S] ,\nonumber\\
\delta _A \Phi & = & -i\{ E_A,S \}
,\;\;\;\;\;\; \delta _A S = i\{
E_A,\Phi \}, \nonumber\\
\delta _V \Phi^\prime & = & [E_V,\Phi^\prime] ,\;\;\;\;\;\;\;\;\;\;
\delta _V S^{\, \prime} = [E_V, S^{\, \prime}],\nonumber \\
\delta _A \Phi^\prime & = & -i\{ E_A, S^{\, \prime} \} +2iS^{\,
\prime} {\rm Tr} E_A ,\nonumber\\
\delta _A S^{\, \prime} & = & i\{
E_A, \Phi^\prime\} -2i \Phi^\prime {\rm Tr} E_A. \label{eq:transva4}
\end{eqnarray}
Since the effective potential term $V_0$ is invariant under the
chiral SU(3)$_L\times$SU(3)$_R$ transformation
implying the invariance under the vector transformation, 
then we have
%\begin{widetext}
\begin{eqnarray}
\delta _V V_0 & = & {\rm Tr} \left[\frac{\partial V_0}{\partial \Phi}
\delta _V \Phi + \frac{\partial V_0}{\partial S} \delta _V S \right] +
\left[ (S,\Phi )\rightarrow (S^{\, \prime} , \Phi^\prime ) \right]
= 0 \, . 
\label{eq:varv01}
\end{eqnarray}
The U(1)$_A$ symmetry is explicitly broken by the anomaly in $V_\eta$, 
so that the axial transformation is expressed as
\begin{eqnarray}
\delta _A V_0 & = & {\rm Tr} \left[ \frac{\partial V_0}{\partial \Phi}
\delta _A \Phi + \frac{\partial V_0}{\partial S} \delta _A S \right] +
\left[ (S,\Phi )\rightarrow (S^{\, \prime} , \Phi^\prime ) \right]
= 
\delta V_\eta
\,, \label{eq:varv0}
\end{eqnarray}
where $\delta V_{\eta}$ arises from the chiral anomaly. Then,
using Eqs.(\ref{eq:transva4}), (\ref{eq:varv01}) and 
(\ref{eq:varv0}) as well as the
arbitrariness of the variations $E_V$ and $E_A$ yields the following generating equations:
\begin{eqnarray}
& & \left[ \Phi , \frac{\partial V_0}{\partial \Phi} \right] +
\left[ S , \frac{\partial V_0}{\partial S} \right] + 
(S,\Phi)\rightarrow (S^{\, \prime} , \Phi^\prime ) =0 \,, \nonumber\\
%%%%%%%%%%%%%%%%%%%%%%%%%%%%%%%%%%%%%%%%%%%%%%%%%%%%%%%%%%%%%%%%%%%%%%%%
& & \left\{ \Phi , \frac{\partial V_0}{\partial S} \right\} -
\left\{ S , \frac{\partial V_0}{\partial \Phi} \right\} +
(S,\Phi)\rightarrow (S^{\prime} , \Phi^\prime ) = {\bf 1} 
\left[2{\rm Tr} \left[\Phi^\prime 
\frac{\partial V_0}{\partial S^{\, \prime}}  - S^{\,\prime} 
\frac{\partial V_0}{\partial \Phi^\prime} \right] 
%- \frac{\delta \mathcal{L}_\eta }{3i\alpha}\right] 
+ \frac{\delta V_\eta }{3i\alpha}\right] 
\,. \label{eq:gge}
\end{eqnarray}

In the following analysis, we use the stationary conditions for the 
potential $V_0$ given by
\begin{eqnarray}
\left\langle \frac{\partial V_0}{\partial S} \right\rangle =0
,\;\;\;\;  \,\,\, \left\langle \frac{\partial V_0}{\partial S^{\,
\prime}} \right\rangle = 0 . \nonumber
\end{eqnarray}
Furthermore, we work in the chiral limit, therefore VEVs of $S$ and 
$S'$ are proportional to the unit matrix:
\begin{equation}
\langle {S_i}^j \rangle = v_2 {\delta_i}^j
\ , \quad
\langle {S'_i}^j \rangle = v_4 {\delta_i}^j\
\ . \nonumber
\end{equation}

Differentiating Eq.~(\ref{eq:gge}) with respect to $\Phi$, we obtain
\begin{eqnarray}
2 v_2 \left\langle
  \frac{\partial^2 V_0}{\partial \Phi_i^j \partial \Phi_k^l}
\right\rangle 
+ 2 v_4 \left\langle
  \frac{\partial^2 V_0}{\partial \Phi_i^j \partial \Phi_k^{\prime \,
  l}}
\right\rangle  
= \delta_l^k \left[ 
2 v_{4} \sum_{m=1}^3 \left\langle
 \frac{\partial^2 V_0}{\partial \Phi_i^j \partial \Phi_m^{\prime \,  m}}
\right\rangle -  \frac{1}{3i\alpha}
  \left\langle \frac{\partial \delta V_\eta }
  {\partial \Phi_i^j}\right\rangle \right] 
\,. %\nonumber
\end{eqnarray}
Differentiation of Eq.~(\ref{eq:gge}) with respect to $\Phi'$ gives
\begin{eqnarray}
2 v_2 \left\langle
  \frac{\partial^2
  V_0}{\partial \Phi_i^{\prime \,j} \partial \Phi_k^l} 
\right\rangle 
+ 2 v_4 \left\langle
  \frac{\partial^2 V_0}
       {\partial \Phi_i^{\prime \, j} \partial \Phi_k^{\prime \, l}} 
\right\rangle  
=
\delta_l^k 
\left[ 2v_{4} \sum_{m=1}^3 \left\langle
    \frac{\partial V_0}
         {\partial \Phi_i^{\prime \, j} \partial \Phi_m^{\prime \,m}}
  \right\rangle 
  - \frac{1}{3i\alpha}
    \left\langle 
      \frac{\partial \delta V_\eta }
           {\partial \Phi_i^{\prime \, j}} 
    \right\rangle \right]
\,.%\nonumber
\end{eqnarray}
%\end{widetext}

Along the same method, differentiating Eq.~(\ref{eq:gge}) twice with respect to $S, S^\prime, \Phi$ or $\Phi^\prime$ and using
\begin{eqnarray}
\partial /\partial\pi^0 & = & (\partial/\partial \Phi_1^1 - \partial/\partial \Phi_2^2)/\sqrt{2} , \nonumber\\
\partial /\partial\pi^{\prime 0} & = & (\partial/\partial \Phi_1^{\prime 1} - \partial/\partial \Phi_2^{\prime 2})/\sqrt{2} , \nonumber\\
\partial /\partial f_a & = & (\partial/\partial S_1^1 + \partial/\partial S_2^2)/\sqrt{2} , \nonumber\\
\partial /\partial f_b & = & \partial/\partial S_3^3 , \nonumber\\
\partial /\partial f_c & = & (\partial/\partial S_1^{\prime 1} + \partial/\partial S_2^{\prime 2})/\sqrt{2} , \nonumber\\
\partial /\partial f_d & = & \partial/\partial S_3^{\prime 3},%\nonumber
\end{eqnarray}
we obtain the relations between three-point and two-point
couplings in the isospin limit as
%\begin{widetext}
\begin{eqnarray}
& & v_2 \left\langle \frac{\partial^3 V_0}{\partial {(\pi^0)}^2
\partial f_a } \right\rangle + v_4 \left\langle \frac{\partial^3
V_0}{\partial \pi^0  \partial \pi^{\prime \, 0} \partial f_a }
\right\rangle = \frac{1}{\sqrt{2}} \left\langle \frac{\partial^2
V_0}{\partial (f_a)^2 } \right\rangle - \frac{1}{\sqrt{2}} \left\langle
\frac{\partial^2 V_0}{ \partial (\pi^0)^2 } \right\rangle ,\nonumber\\
%%%%%%%%%%%%%%%%%%%%%%%%%%%%%%%%%%%%%%%%%%%%%%%%%%%%%%%%%%%%%%%%%%%%%%%%
& & v_2 \left\langle \frac{\partial^3 V_0}{\partial {(\pi^0)}^2
\partial f_b } \right\rangle + v_4 \left\langle \frac{\partial^3
V_0}{\partial \pi^0 \partial \pi^{\prime \, 0} \partial f_b }
\right\rangle = \frac{1}{\sqrt{2}} \left\langle \frac{\partial^2
V_0}{ \partial f_a \partial f_b } \right\rangle ,\nonumber\\
%%%%%%%%%%%%%%%%%%%%%%%%%%%%%%%%%%%%%%%%%%%%%%%%%%%%%%%%%%%%%%%%%%%%%%%%
& &  v_2 \left\langle \frac{\partial^3 V_0}{ \partial \pi^0
\partial \pi^{\prime \, 0} \partial f_{a} } \right\rangle + v_4 \left\langle
\frac{\partial^3 V_0}{\partial {(\pi^{\prime \, 0})}^2  \partial f_{a} }
\right\rangle = \frac{1}{\sqrt{2}} \left\langle \frac{\partial^2 V_0}{\partial f_{a} \partial f_c } \right\rangle 
- \frac{1}{\sqrt{2}} \left\langle \frac{\partial^2 V_0}{ \partial
\pi^0 \partial \pi^{\prime \, 0} } \right\rangle ,\nonumber\\
%%%%%%%%%%%%%%%%%%%%%%%%%%%%%%%%%%%%%%%%%%%%%%%%%%%%%%%%%%%%%%%%%%%%%%%%%%%%%%%%
& & v_2 \left\langle \frac{\partial^3 V_0}{ \partial \pi^0
\partial \pi^{\prime \, 0} \partial f_b } \right\rangle + v_4 \left\langle
\frac{\partial^3 V_0}{\partial {(\pi^{\prime \, 0})}^2  \partial f_b }
\right\rangle = \frac{1}{\sqrt{2}}  \left\langle \frac{\partial^2 V_0}{\partial f_b \partial f_c } \right\rangle ,\nonumber\\
%%%%%%%%%%%%%%%%%%%%%%%%%%%%%%%%%%%%%%%%%%%%%%%%%%%%%%%%%%%%%%%%%%%%%%%%%%%%%%%%
& & v_2 \left\langle \frac{\partial^3 V_0}{\partial
{(\pi^0)}^2 \partial f_c } \right\rangle + v_4 \left\langle \frac{\partial^3 V_0}{\partial \pi^0 \partial \pi^{\prime \, 0} \partial f_c } \right\rangle = \frac{1}{\sqrt{2}}
 \left\langle \frac{\partial^2 V_0}{ \partial f_a \partial f_c } \right\rangle
- \frac{1}{\sqrt{2}} \left\langle  \frac{\partial^2 V_0}{ \partial
\pi^0 \partial \pi^{\prime \, 0}} \right\rangle ,\nonumber\\
%%%%%%%%%%%%%%%%%%%%%%%%%%%%%%%%%%%%%%%%%%%%%%%%%%%%%%%%%%%%%%%%%%%%%%%%%%%%%%%%
& & v_2 \left\langle \frac{\partial^3 V_0}{\partial {(\pi^0)}^2
\partial f_d } \right\rangle + v_4 \left\langle \frac{\partial^3
V_0}{\partial \pi^0 \partial \pi^{\prime \, 0} \partial f_d }
\right\rangle = \frac{1}{\sqrt{2}} \left\langle \frac{\partial^2 V_0}{ \partial f_a \partial f_d } \right\rangle ,\nonumber\\
%%%%%%%%%%%%%%%%%%%%%%%%%%%%%%%%%%%%%%%%%%%%%%%%%%%%%%%%%%%%%%%%%%%%%%%%%%%%%%%%
& & v_2 \left\langle \frac{\partial^3 V_0}{ \partial \pi^0
\partial \pi^{\prime \, 0}  \partial f_c } \right\rangle +
 v_4 \left\langle \frac{\partial^3 V_0}{\partial {(\pi^{\prime \, 0})}^2
\partial f_c } \right\rangle =  \frac{1}{\sqrt{2}}
\left\langle \frac{\partial^2 V_0}{\partial (f_c)^2 }
\right\rangle - \frac{1}{\sqrt{2}}
\left\langle \frac{\partial^2 V_0}{ \partial (\pi^{\prime \, 0})^2 } \right\rangle ,\nonumber\\
%%%%%%%%%%%%%%%%%%%%%%%%%%%%%%%%%%%%%%%%%%%%%%%%%%%%%%%%%%%%%%%%%%%%%%%%%%%%%%%%
& & v_2 \left\langle \frac{\partial^3 V_0}{ \partial \pi^0
\partial \pi^{\prime \, 0}  \partial f_d } \right\rangle + v_4 \left\langle
\frac{\partial^3 V_0}{ \partial {(\pi^{\prime \, 0})}^2 \partial f_d }
\right\rangle = \frac{1}{\sqrt{2}} \left\langle \frac{\partial^2 V_0}{ \partial f_c \partial f_d } \right\rangle . %\nonumber
\end{eqnarray}
%\end{widetext}

And, differentiating Eq.~(\ref{eq:gge}) three times with respect to $S, S^\prime, \Phi$ or $\Phi^\prime$, the relations between four-point and three-point coupling in the
isospin limit are obtained as 
%\begin{widetext}
\begin{eqnarray}
& & v_2 \left\langle \frac{\partial^4 V_0}{\partial (\pi^0)^4 } \right\rangle + v_4 \left\langle \frac{\partial^4
V_0}{\partial (\pi^0)^3 \partial \pi^{\prime \, 0}} \right\rangle  = 3
\frac{1}{\sqrt{2}} \left\langle \frac{\partial^3 V_0}{
\partial (\pi^0)^2 \partial f_a} \right\rangle , \nonumber\\
%%%%%%%%%%%%%%%%%%%%%%%%%%%%%%%%%%%%%%%%%%%%%%%%%%%%%%%%%%%%%%%%%%%%%%%%%%%%%%%%%%%%%%%%%%
& & v_2 \left\langle \frac{\partial^4 V_0}{\partial \pi^{\prime \, 0} \partial {(\pi^0)}^{3}} \right\rangle + v_4 \left\langle \frac{\partial^4 V_0}{\partial \pi^{\prime \, 0}
\partial {(\pi^0)}^{2} \partial \pi^{\prime \, 0}} \right\rangle = 2 \frac{1}{\sqrt{2}} \left\langle \frac{\partial^3
V_0}{\partial \pi^{\prime \, 0} \partial \pi^0 \partial f_a } \right\rangle + \frac{1}{\sqrt{2}} \left\langle \frac{\partial^3 V_0}{ \partial {(\pi^0)}^2  \partial f_c }  \right\rangle ,\nonumber\\
%%%%%%%%%%%%%%%%%%%%%%%%%%%%%%%%%%%%%%%%%%%%%%%%%%%%%%%%%%%%%%%%%%%%%%%%%%%%%%%%%%%%%%%%%%
& & v_2 \left\langle \frac{\partial^4 V_0}{\partial \pi^0
\partial {(\pi^{\prime \, 0})}^2 \partial \pi^0} \right\rangle + v_4
\left\langle \frac{\partial^4 V_0}{\partial \pi^0 \partial {(\pi^{\prime \, 0})}^3 } \right\rangle 
= \frac{1}{\sqrt{2}} \left\langle \frac{\partial^3 V_0}{
\partial {(\pi^{\prime \, 0})}^{2} \partial f_a } \right\rangle + 2
\frac{1}{\sqrt{2}} \left\langle \frac{\partial^3 V_0}{\partial \pi^0
\partial \pi^{\prime \, 0} \partial f_c } \right\rangle , \nonumber\\
%%%%%%%%%%%%%%%%%%%%%%%%%%%%%%%%%%%%%%%%%%%%%%%%%%%%%%%%%%%%%%%%%%%%%%%%%%%%%%%%%%%%%%%%%%
& & v_2 \left\langle \frac{\partial^4 V_0}{\partial {(\pi^{\prime \, 0})}^3 \partial \pi^0} \right\rangle + v_4 \left\langle \frac{\partial^4
V_0}{\partial {(\pi^{\prime \, 0})}^4}
\right\rangle = 3 \frac{1}{\sqrt{2}} \left\langle \frac{\partial^3
V_0}{ \partial {(\pi^{\prime \, 0})}^2 \partial f_c } \right\rangle .%\nonumber
\end{eqnarray}
%\end{widetext}

The above relations are re-expressed in terms of the mixing angles
for the scalar and pseudoscalar mesons in Eqs.~(\ref{eq:mixings}) and (\ref{eq:mixingp}),
the pion decay constant and the VEVs of $M$ and $M'$ fields as
\begin{eqnarray}
& & v_2 = \frac{1}{2} \left[ F_\pi \cos \theta_\pi + \tilde
F_\pi \sin \theta_\pi \right] = \frac{1}{2} \sum_{j=1}^2 (F_\pi)_j
(U_\pi^{-1})_{aj} , \nonumber\\
%%%%%%%%%%%%%%%%%%%%%%%%%%%%%%%%%%%%%%%%%%%%%%%%%%%%%%%%%%%%%%%%%%%%%%%%
& & v_4 = \frac{1}{2} \left[ - F_\pi \sin \theta_\pi + \tilde
F_\pi \cos \theta_\pi \right] = \frac{1}{2} \sum_{j=1}^2 (F_\pi)_j
(U_\pi^{-1})_{bj} ,
\end{eqnarray}
where $(F_\pi)_1 = F_\pi, (F_\pi)_2 = \tilde{F}_\pi$ are the decay constants of $\pi(140)$ and $\pi(1300)$, respectively, and the mixing matrix $U_\pi$ is defined by 
\begin{eqnarray}
U_\pi & = & \left(
\begin{array}{lr}
(U_\pi)_{1a} & (U_\pi)_{1b} \\
(U_\pi)_{2a} & (U_\pi)_{2b} \\
\end{array}
\right) = \left(
\begin{array}{lr}
\cos \theta _\pi & -\sin \theta _\pi \\
\sin \theta _\pi & \cos \theta _\pi \\
\end{array}
\right)
.\nonumber
\end{eqnarray}
Note that, in the chiral limit, only the NG boson can couple to
the axial vector current, so that $\tilde F_\pi =0$. 
Then, writing the mass matrices as
\begin{eqnarray}
(\tilde{m}^2_\pi)_{AB} \equiv \left\langle \frac{\partial^2 V}{\partial
\pi_A\partial \pi_B } \right\rangle , \;\;\;\;  (\tilde{m}^2_f)_{CD}
\equiv \left\langle \frac{\partial^2 V}{\partial f_C \partial f_D }
\right\rangle,\nonumber 
\end{eqnarray}
where subscripts $A$ and $B$ run from $a$ to $b$ and $C$ and $D$ from $a$ to $d$, we have
%\begin{widetext}
\begin{eqnarray}
& & \frac{1}{\sqrt{2}} F_\pi \sum_{A=a}^b (U_\pi)_{1A} \left\langle
\frac{\partial^3 V_0}{\partial (\pi^0)_a  \partial (\pi^0)_A
\partial f_a } \right\rangle
= (\tilde{m}^2_f)_{aa} - (\tilde{m}^2_\pi)_{aa} , \nonumber\\
%%%%%%%%%%%%%%%%%%%%%%%%%%%%%%%%%%%%%%%%%%%%%%%%%%%%%%%%%%%%%%%%%%%%%%%%%%%%%%%%
& & \frac{1}{\sqrt{2}} F_\pi \sum_{A=a}^b (U_\pi)_{1A} \left\langle
\frac{\partial^3 V_0}{\partial (\pi^0)_a  \partial (\pi^0)_A
\partial f_b } \right\rangle = (\tilde{m}^2_f)_{ab} ,  \nonumber\\
%%%%%%%%%%%%%%%%%%%%%%%%%%%%%%%%%%%%%%%%%%%%%%%%%%%%%%%%%%%%%%%%%%%%%%%%%%%%%%%
& & \frac{1}{\sqrt{2}} F_\pi \sum_{A=a}^{b} (U_\pi)_{1A}
\left\langle \frac{\partial^3 V_0}{ \partial (\pi^0)_A \partial
(\pi^0)_b \partial f_{a} } \right\rangle = (\tilde{m}^2_f)_{ac} - (\tilde{m}^2_\pi)_{ab}  \nonumber\\
%%%%%%%%%%%%%%%%%%%%%%%%%%%%%%%%%%%%%%%%%%%%%%%%%%%%%%%%%%%%%%%%%%%%%%%%%%%%%%%
& & \frac{1}{\sqrt{2}} F_\pi \sum_{A=a}^{b} (U_\pi)_{1A}
\left\langle \frac{\partial^3 V_0}{ \partial (\pi^0)_A \partial
(\pi^0)_b \partial f_{b} } \right\rangle = (\tilde{m}^2_f)_{bc} ,  \nonumber \\
%%%%%%%%%%%%%%%%%%%%%%%%%%%%%%%%%%%%%%%%%%%%%%%%%%%%%%%%%%%%%%%%%%%%%%%%%%%%%%%%
& & \frac{1}{\sqrt{2}} F_\pi \sum_{A=a}^b (U_\pi)_{1A} \left\langle
\frac{\partial^3 V_0}{\partial (\pi^0)_a  \partial (\pi^0)_A
\partial f_c } \right\rangle = (\tilde{m}^2_f)_{ac} - (\tilde{m}^2_\pi)_{ab} ,  \nonumber\\
%%%%%%%%%%%%%%%%%%%%%%%%%%%%%%%%%%%%%%%%%%%%%%%%%%%%%%%%%%%%%%%%%%%%%%%%%%%%%%%%
& & \frac{1}{\sqrt{2}} F_\pi \sum_{A=a}^b (U_\pi)_{1A} \left\langle
\frac{\partial^3 V_0}{\partial (\pi^0)_a  \partial (\pi^0)_A
\partial f_d } \right\rangle = (\tilde{m}^2_f)_{ad} ,  \nonumber\\
%%%%%%%%%%%%%%%%%%%%%%%%%%%%%%%%%%%%%%%%%%%%%%%%%%%%%%%%%%%%%%%%%%%%%%%%%%%%%%%%
& & \frac{1}{\sqrt{2}} F_\pi \sum_{A=a}^b (U_\pi)_{1A} \left\langle
\frac{\partial^3 V_0}{\partial (\pi^0)_b  \partial (\pi^0)_A
\partial f_c } \right\rangle = (\tilde{m}^2_f)_{cc} - (\tilde{m}^2_\pi)_{bb} ,  \nonumber\\
%%%%%%%%%%%%%%%%%%%%%%%%%%%%%%%%%%%%%%%%%%%%%%%%%%%%%%%%%%%%%%%%%%%%%%%%%%%%%%%%
& & \frac{1}{\sqrt{2}} F_\pi \sum_{A=a}^b (U_\pi)_{1A} \left\langle
\frac{\partial^3 V_0}{\partial (\pi^0)_b  \partial (\pi^0)_A
\partial f_d } \right\rangle = (\tilde{m}_f^2)_{cd},%\nonumber
\end{eqnarray}
%\end{widetext}

Summing these equations with mixing matrices $U_\pi$ and $U_f$
we get the following physical coupling
%\begin{widetext}
\begin{eqnarray}
\frac{1}{\sqrt{2}} F_\pi \left\langle \frac{\partial^3 V_0}{\partial
(\pi^0_p)_1  \partial (\pi^0_p)_k \partial f_j } \right\rangle & = &
\frac{1}{\sqrt{2}} F_\pi \sum_{A,B=a,b} \, \sum_{C=a,b,c,d}
(U_\pi)_{1A} (U_\pi)_{kB} (U_f)_{jC} \left\langle \frac{\partial^3
V_0}{\partial (\pi^0)_B  \partial (\pi^0)_A \partial f_C }
\right\rangle \nonumber\\
& & = \sum_{C=a}^d (U_f)_{jC} \left[ (U_\pi)_{ka} (m^2_f)_{aC} +
(U_\pi)_{kb} (m^2_f)_{cC} \right] \nonumber\\
& &\quad\quad - \sum_{A=a}^{b} (U_\pi)_{kA} \left[ (U_f)_{ja}
(m^2_\pi)_{aA} + (U_f)_{jc} (m^2_\pi)_{Ab} \right] ,
%\nonumber
\end{eqnarray}
where we write the physical pseudoscalars as $(\pi_p^0)_1 \equiv \pi_p^0$ and $(\pi_p^{\prime 0})_2 \equiv \pi_p^{\prime 0}$. 

So that, in the case that the pseudoscalars are neutral, we have
\begin{eqnarray}
g_{f_j \pi\pi} & \equiv & \left\langle \frac{ \partial^3 V}{ \partial
(\pi^0_p)_1
\partial (\pi^0_p)_1 \partial f_j } \right\rangle = \sum_{A=a}^{b} \sum_{B=a}^{b} \sum_{C=a}^{d} (U_\pi)_{1A}
(U_\pi)_{1B} (U_f)_{jC} \left\langle \frac{ \partial^3 V_0}{
\partial (\pi^0)_A \partial (\pi^0)_B \partial f_C } \right\rangle .
\end{eqnarray}
As a result
\begin{eqnarray}
g_{f_j\pi\pi}^2 & = & \left[ \sum_{A,B=a,b}\sum_{C=a,b,c,d}
(U_\pi)_{1A} (U_\pi)_{1B} (U_f)_{jC} \left\langle \frac{\partial^3
V_0}{\partial (\pi^0)_B  \partial (\pi^0)_A \partial f_C }
\right\rangle \right]^2 \nonumber\\
& = & \frac{2}{F_\pi^2} \left( m^2_{f_j} \right)^2 \left[
(U_\pi)_{1a} (U_f)_{ja} + (U_\pi)_{1b} (U_f)_{jc} \right]^2 .\nonumber\\ 
\label{eq:g00j2u}
\end{eqnarray}

Here we have considered that in the chiral limit, the light pions
are massless. Therefore, from this equation, we obtain the following relations 
\begin{eqnarray}
\sum_{i=1}^{4} \dfrac{g_{f_j \pi\pi}^2}{\left( m_{f_j}^2 \right)^2}
& = & \sum_{i=1}^{4} \frac{2}{F_\pi^2}  \left[ (U_\pi)_{1a} (U_f)_{ja} + (U_\pi)_{1b} (U_f)_{jc} \right]^2
= \dfrac{2}{F^2_\pi} , \nonumber\\
% % % % % % % % % % % % % % % % % % % % % % % % % % % % % % % % % % % % % % % %
\sum_{i=1}^{4} \dfrac{g_{f_j \pi\pi}}{m_{f_j}^2} \left( U_f^{-1} \right)_{aj} 
& = & \sum_{i=1}^{4} \frac{\sqrt{2}}{F_\pi} \left[
(U_\pi)_{1a} (U_f)_{ja} + (U_\pi)_{1b} (U_f)_{jc} \right] \left( U_f^{-1} \right)_{aj}
= \dfrac{\sqrt{2}}{F_\pi} \cos \theta_{\pi} .%\nonumber
\end{eqnarray}

Similarly, in the the chiral limit, we can derive that 4-$\pi^0$ coupling constant satisfies
\begin{eqnarray}
g_{\pi \pi \pi \pi} & = & \left\langle \frac{ \partial^4 V}{
\partial (\pi_p^0)_1 \partial (\pi_p^0)_1 \partial (\pi_p^0)_1 \partial (\pi_p^0)_1 }
\right\rangle = \frac{6}{ F_\pi^2 } \sum_{j=1}^{4} m_{f_j}^2
\Bigg[ (U_\pi)_{1a} (U_f)_{ja} + (U_\pi)_{1b} (U_f)_{jc} \Bigg]^2.
\label{eq:fourpoint_h}
\end{eqnarray}
which yields 
\begin{eqnarray}
%%%%%%%%%%%%%%%%%%%%%%%%%%%%%%%%%%%%%%%%%%%%%%%%%%%%%%%%%%%%%%%%%
\sum_{i=1}^{4}
\dfrac{g_{f_j\pi\pi}^2}{m_{f_j}^2} & = & \dfrac{1}{3} g_{\pi \pi \pi \pi},
\end{eqnarray}
by using Eqs.(\ref{eq:g00j2u}, \ref{eq:fourpoint_h}).
%\end{widetext}

\section{Reduction of the matrix element (\ref{eq:decaymatrix})}

\label{app:reduction}

%\begin{widetext}
Here, we provide the reduction of the heavy meson decay matrix element (\ref{eq:decaymatrix}). Its first term can be rewritten as
\begin{eqnarray}
\sum_{i = 1}^{4} g_{f_i\pi\pi} \frac{(p_{\pi_1} + p_{\pi_2})^\mu}{s
- m_{f_i}^2 + im_{f_i}\Gamma_{f_i}(s)} (U_f^{-1})_{ai} & = &
g_{f_1\pi\pi} \frac{(p_{\pi_1} + p_{\pi_2})^\mu}{s - m_{f_1}^2 +
im_{f_1}\Gamma_{f_1}(s)} (U_f^{-1})_{a1}\nonumber\\
& & {} + \sum_{i = 2}^{4} g_{f_i\pi\pi} \frac{(p_{\pi_1} +
p_{\pi_2})^\mu}{s - m_{f_i}^2 + im_{f_i}\Gamma_{f_i}(s)}
(U_f^{-1})_{ai}.\nonumber
\end{eqnarray}
%\end{widetext}
Considering that the momentum carried by the exchanged scalar meson
is smaller than the heavier scalar meson mass, i.e., $s < m_{f_i}^2,
(i = 2,3,4)$, we make the expansion
%\begin{widetext}
\begin{eqnarray}
\sum_{i = 2}^{4} g_{f_i\pi\pi} \frac{(p_{\pi_1} + p_{\pi_2})^\mu}{s
- m_{f_i}^2 + im_{f_i}\Gamma_{f_i}(s)} (U_f^{-1})_{ai} & = & - \sum_{i =
2}^{4} g_{f_i\pi\pi} \frac{(p_{\pi_1} + p_{\pi_2})^\mu}{m_{f_i}^2 -
im_{f_i}\Gamma_{f_i}(s)} (U_f^{-1})_{ai} \nonumber\\
& & \times \left[1 + \frac{s}{m_{f_i}^2 + im_{f_i}\Gamma_{f_i}(s)} +
\left(\frac{s}{m_{f_i}^2 + im_{f_i}\Gamma_{f_i}(s)}\right)^2 +
\cdots\right].\nonumber
\end{eqnarray}
And also concerning that $\Gamma_{f_i} \ll m_{f_{i}}, (i = 2,3,4)$, we can
simplify the above relation as
\begin{eqnarray}
\sum_{i = 2}^{4} g_{f_i\pi\pi} \frac{(p_{\pi_1} + p_{\pi_2})^\mu}{s
- m_{f_i}^2 + im_{f_i}\Gamma_{f_i}(s)} (U_f^{-1})_{ai} & = & - \sum_{i =
2}^{4} g_{f_i\pi\pi} \frac{(p_{\pi_1} +
p_{\pi_2})^\mu}{m_{f_i}^2} (U_f^{-1})_{ai} \left[1 + \frac{s}{m_{f_i}^2} +
\left(\frac{s}{m_{f_i}^2}\right)^2 + \cdots\right].\nonumber
\end{eqnarray}

With respect to the relation
\begin{eqnarray}
\sum_{i = 1}^{4} \frac{g_{f_i\pi\pi}}{m_{f_i}^2} (U_f^{-1})_{ai} & = &
\frac{\sqrt{2}}{F_\pi}\cos\theta_\pi,%\nonumber
\end{eqnarray}
and neglecting the terms of and higher than $\mathcal{O}(s/m_{f_i}^2),
(i =2,3,4)$ we obtain
\begin{eqnarray}
\sum_{i = 2}^{4} g_{f_i\pi\pi} \frac{(p_{\pi_1} + p_{\pi_2})^\mu}{s
- m_{f_i}^2 + im_{f_i}\Gamma_{f_i}(s)} (U_f^{-1})_{ai} & = & -
(p_{\pi_1} +
p_{\pi_2})^\mu\left(\frac{\sqrt{2}}{F_\pi}\cos\theta_\pi -
\frac{g_{\sigma\pi\pi}}{m_{\sigma}^2} (U_f^{-1})_{a1} \right). %\nonumber
\end{eqnarray}
Then, we finally arrive at the sum rule for the scalar meson
exchanging contribution
\begin{eqnarray}
\sum_{i = 1}^{4} g_{f_i\pi\pi} \frac{(p_{\pi_1} + p_{\pi_2})^\mu}{s
- m_{f_i}^2 + im_{f_i}\Gamma_{f_i}(s)} (U_f^{-1})_{ai} & = &
g_{f_1\pi\pi} \frac{(p_{\pi_1} + p_{\pi_2})^\mu}{s - m_{f_1}^2 +
im_{f_1}\Gamma_{f_1}(s)} (U_f^{-1})_{a1} \nonumber\\
& & {} - (p_{\pi_1} +
p_{\pi_2})^\mu\left(\frac{\sqrt{2}}{F_\pi}\cos\theta_\pi -
g_{\sigma\pi\pi}
\frac{1}{m_{\sigma}^2}(U_f^{-1})_{a1}\right). %\nonumber
\end{eqnarray}
It should be noticed that the second term of this sum rule arising
from the heavier resonance contribution and only the first term
leaves once the heavier resonances are neglected from the beginning.
\vspace{0.5cm}
\end{widetext}

%%%%%%%%%%%%%%%%%%%%%%%%%%%%%%%%%%%%%%%%%%%%%%%%%%%%%%%%%%%%%%%%%%%%%%%%%%%%%%%%%%

\end{document}